\documentclass[10pt,twocolumn,twoside]{IEEEtran}
\usepackage{amsbsy,amsmath,amsfonts,amssymb,amsbsy,subfigure,booktabs,multirow}
\usepackage{bm,cite,graphicx,psfrag,pstricks,theorem,tikz,times,url,verbatim}
\usetikzlibrary{shapes,snakes,calendar,matrix,backgrounds,folding}
\usepackage[english]{babel}
\usetikzlibrary{arrows,automata}
\usetikzlibrary{calc} 
\usetikzlibrary{decorations.pathmorphing} 

\newcommand{\bi}{\begin{itemize}}
\newcommand{\ei}{\end{itemize}}
\newcommand{\ben}{\begin{enumerate}}
\newcommand{\een}{\end{enumerate}}
\newcommand{\bc}{\begin{cases}}
\newcommand{\ec}{\end{cases}}
\newcommand{\bd}{\begin{description}}
\newcommand{\ed}{\end{description}}

\newcommand{\be}{\begin{equation}}
\newcommand{\ee}{\end{equation}}
\newcommand{\bea}{\begin{eqnarray}}
\newcommand{\eea}{\end{eqnarray}}

\newcommand\T{\rule{0pt}{2.0ex}}
\newcommand\B{\rule[-0.8ex]{0pt}{0pt}}

\newtheorem{thm}{Theorem}

\newtheorem{propos}{Proposition}
\newtheorem{corol}{Corollary}

\newtheorem{algo}{Algorithm}

\theoremstyle{plain}
\newtheorem{remark}{Remark}

\newcommand{\nn}{\nonumber}

\graphicspath{%
     {./Figures/}
}

\begin{document}

\title{
Cross-layer design of distributed sensing-estimation with quality feedback, Part II: Myopic schemes
}

\author{Nicol\`{o}~Michelusi~and~Urbashi~Mitra

\thanks{Copyright (c) 2014 IEEE. Personal use of this material is permitted. However, permission to use this material for any other purposes must be obtained from the IEEE by sending a request to pubs-permissions@ieee.org.}
\thanks{N. Michelusi and U. Mitra are with the Department of Electrical Engineering, University of Southern California. email addresses: \{michelus,ubli\}@usc.edu.}
\thanks{This research has been funded in part by the following grants:
ONR N00014-09-1-0700, CCF-0917343, CCF-1117896, CNS-1213128, AFOSR FA9550-12-1-0215, and DOT CA-26-7084-00.
N. Michelusi is in part supported by AEIT (Italian association of electrical engineering) through the scholarship "Isabella Sassi Bonadonna 2013".
}
\thanks{Parts of this work have appeared in \cite{MicheAllerton,MicheGlobalsip}.}
}

\maketitle
\noindent\begin{abstract}
 This two-part paper presents a feedback-based cross-layer framework for distributed sensing and estimation of a dynamic process by a wireless sensor network (WSN).  Sensor nodes wirelessly communicate measurements to the fusion center (FC). Cross-layer factors such as packet collisions and the sensing-transmission costs are considered. Each SN adapts its sensing-transmission action based on its own local observation quality and the estimation quality feedback from the FC under cost constraints for each SN. In this second part, low-complexity \emph{myopic} sensing-transmission policies (MPs) are designed to optimize a trade-off between performance and the cost incurred by each SN. The MP is computed in closed form for a \emph{coordinated} scheme, whereas an iterative algorithm is presented for a \emph{decentralized} one, which converges to a local optimum. The MP dictates that, when the estimation quality is poor, only the \emph{best} SNs activate, otherwise all SNs remain idle to preserve energy. For both schemes, the threshold on the estimation quality below which the SNs remain idle is derived in closed form, and is shown to be independent of the number of channels. It is also proved that a single channel suffices for severely energy constrained WSNs. The proposed MPs are shown to yield near-optimal performance with respect to the optimal policy of Part~I~\cite{MichelusiP1}, at a fraction of the complexity, thus being more suitable for practical WSN deployments.
\end{abstract}
\vspace{-3mm}
\section{Introduction}
 Wireless sensor networks (WSNs) enable the monitoring of large areas via many low powered sensor nodes (SNs)
with data acquisition, processing and communication capabilities \cite{Romer}.
However, WSN design is challenged by
the high optimization complexity typical of multi-agent systems~\cite{Bernstein},
necessitating decentralized SN operation based on \emph{local} information and limited feedback,
and needs to explicitly consider the resource constraints of SNs.

In this two part paper, we present a {feedback-based} cross-layer framework for distributed sensing and estimation of a time-correlated random process at a fusion center (FC),
based on noisy measurements collected from nearby SNs,
which accounts for cross-layer factors such as the shared wireless channel, resulting in collisions among SNs, the sensing and transmission costs,
and the \emph{local} state and local view of the SNs.
In order to cope with the uncertainties and stochastic dynamics introduced by these cross-layer components,
the FC broadcasts feedback information to the SNs, based on the estimation quality achieved, thus enabling 
adaptation of their sensing-transmission action.
We design joint sensing-transmission policies with
the goal to minimize the mean squared estimation error (MSE) at the FC,
under a constraint on the sensing-transmission cost incurred by each SN.

In Part I, 
we provided a theoretical foundation for the reduction of the system complexity,
arising from the local asymmetries
 due to the decentralized operation of SNs, their local state and local view, and the multi-agent nature of the system,
 by exploiting the \emph{statistical symmetry} of the WSN with respect to the local view of the SNs and the \emph{large network approximation}.
However, the
dynamic programming (DP) algorithms designed in Part I
still have high complexity. In this second part, building on the results derived in Part I, 
we design low-complexity \emph{myopic policies} for a \emph{coordinated scheme},
where the FC schedules the action (sense and transmit, or remain idle) of each SN, and 
a \emph{decentralized scheme}, where the SNs determine their action in a decentralized fashion,
based on the feedback information and on their local accuracy state.
These myopic policies are designed in such a way as to optimize a trade-off between the MSE at the FC and the
sensing-transmission cost incurred by each SN.

For the coordinated scheme, we derive the myopic policy in closed form. For the decentralized scheme,
we present an iterative algorithm based on the bisection method \cite{bisection}, which converges provably to a local optimum of the myopic cost function.
Similar to the optimal policy derived via DP, the myopic policy dictates that, when the estimation quality at the FC is poor, the 
SNs with the best observation quality activate  by collecting high accuracy measurements and transmit them to the FC, to improve the estimation quality.
In contrast, if the estimation quality is good, the SNs stay idle to preserve energy. For both schemes, we derive, in closed form,  the value of the threshold on the estimation quality below which the SNs remain idle, and show that it is independent of the number of channels $B$ employed. Additionally, we prove that, for severely energy constrained systems, one orthogonal channel ($B{=}1$) suffices. Numerically, we show that  the myopic policies achieve near-optimal performance with respect to the globally optimal DP policy, at a fraction of the complexity, and are thus suitable for implementation in practical WSN deployments.

The problem of decentralized estimation and detection has seen a vast research effort in the last decade,
especially in the design of optimal schemes 
for parameter estimation \cite{Xiao,Thatte,Xiao2}, hypothesis testing \cite{Ray,Tsitsiklis,Chamberland}, tracking \cite{Saber,Epstein} and random field estimation \cite{Fang}.
Distributed estimation in bandwidth-energy constrained environments has been considered in \cite{Chieh,Ribeiro,Msechu,Junlin}, for a static setting.
 Estimation and detection problems exploiting
feedback information from the FC have been investigated in \cite{Dogandzic,Peng,Kreidl,Dey},
\emph{e.g.}, enabling adaptation of the SNs' quantizers in the estimation of a finite state Markov chain \cite{Dey}.
A consensus based approach for distributed multi-hypothesis testing has been studied in~\cite{Saligrama}.

Differently from these works, we employ a cross-layer perspective, \emph{i.e.}, we jointly consider and optimize the resource constraints typical of WSNs,
such as the shared wireless channel, resulting in collisions among SNs, the time-varying sensing capability of the SNs, their decentralized decisions,
and the cost of sensing and data transmission, and propose a feedback mechanism from the FC to enable \emph{adaptation} and cope with the random fluctuations  in the overall measurement quality collected at the FC,
induced by these cross-layer factors.
This is in contrast to, \emph{e.g.}, \cite{Dey}, where adaptation serves to cope with the distortion introduced by quantization.
We do not consider the problem of quantizer design,
and focus instead on a  \emph{censoring} approach \cite{Appadwedula,Msechu},
 \emph{i.e.}, quantization is fixed and sufficiently fine-grained, so that the measurements received at the FC can be approximated as Gaussian.
In fact, in light of our cross-layer design perspective, quantization may be less relevant due to the overhead required to perform essential tasks such as
synchronization and channel estimation~\cite{Appadwedula}.

Distributed Kalman filtering for WSNs has been proposed in \cite{Olfati},
using a consensus approach and local Kalman filters at each SN. 
In this paper, Kalman filtering is employed only at the FC, which collects unfiltered observations from the SNs.
In fact, due to the poor estimation capability of SNs and their energy constraints, which force them to 
 remain idle most of the time,
the performance gain achievable by exploiting the time-correlation via local Kalman filtering may be small.

 This paper is organized as follows.
  In Secs. \ref{model}, we present the system model
and some preliminary results from Part~I.
  In Secs. \ref{analysisCoord} and \ref{analysisDistr}, we derive the myopic policy for the coordinated and decentralized schemes, respectively.
In Sec.~\ref{numres}, we provide numerical results.
In Sec. \ref{conclusions}, we conclude the paper.
The analytical proofs are provided in the Appendix.

\vspace{-3mm}
\section{System Model}
\label{model}

 \begin{table*}[t]
\caption{Main system parameters}
\label{tab1}
\vspace{-5mm}
\begin{center}
\footnotesize
\scalebox{0.88}{
\begin{tabular}{|c| l | c | l | c | l | c | l |}
\hline\T\B $\{X_k\}$& random process to be tracked
&
$S_A$& local ambient SNR
 &
$Y_{n,k}$& measurement of SN $n$ in slot $k$
&
$\gamma_{n,k}$& accuracy state with 
 s.s.d. $\pi_{\gamma}(\gamma)$
  \\\hline\T\B 
  $\alpha$& time-correlation parameter
&$S_{M,n,k}$& local measurement SNR
&$A_{n,k}$& activation of SN $n$, slot $k$
&$B_{n,k}$& channel ID for SN $n$, slot $k$
\\\hline\T\B
 $\Lambda_k$& aggregate SNR at FC
& $\phi S_{M,n,k}$& sensing cost
& $c_{\mathrm{TX}}$& transmission cost
 &$B$& \# channels available, $B\leq N_S$
 \\\hline\T\B
$V_k$ & prior variance 
&$\hat V_k$ & posterior variance 
&$q$ & SN activation  probability 
& $N_S$ & \# of SNs, $N_S\geq B$
\\\hline\T\B 
$\theta{\triangleq}\frac{\phi}{c_{\mathrm{TX}}}$& normalized unitary sensing cost &
$\bar M_{\delta}$ & average MSE
&
$\bar C_{\delta}^{n}$ & 
\multicolumn{3}{l|}{
average sensing-transmission cost of SN $n$
}
\\
\hline
\end{tabular}
}
\end{center}
\vspace{-5mm}
\end{table*}

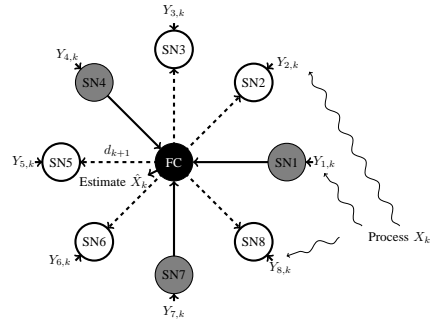
\begin{figure}
    \centering
\scalebox{0.5}{
\begin{tikzpicture}
\draw [ultra thick, ->] (0,0) -- (-0.7,-0.35);
\node at (-1.6,-.5) {Estimate $\hat X_{k}$};
\draw [ultra thick, ->] (3,0) -- (0+0.5,0);
\draw [ultra thick, ->] (-2.12,2.12) -- (0-0.3535,0+0.3535);
\draw [ultra thick, ->] (0,-3) -- (0,0-0.5);
\node at (6,-2) {Process $X_k$};
\draw [fill=black,draw=black,thick,text=white] (0,0) circle [radius=0.5];
\draw [fill=gray,draw=black,thick,text=white] (3,0) circle [radius=0.5];
\node at (3,0) {SN1};
\node at (4,0) {$Y_{1,k}$};
\draw [ultra thick, ->] (3.7,0) -- (3.5,0);
\def\x{3.5}
\def\y{0.35}
\draw [fill=white, ultra thick] (2.12,2.12) circle [radius=0.5];
\node at (2.12,2.12) {SN2};
\node at (2.83+0.2,2.83-0.2) {$Y_{2,k}$};
\draw [ultra thick,dashed, ->] (0,0) -- (2.47-0.7,2.47-0.7);
\draw [ultra thick, ->] (2.62,2.62) -- (2.47,2.47);
\def\x{2.12+0.6}
\def\y{2.12-0.25}
\draw [fill=white, ultra thick] (0,3) circle [radius=0.5];
\node at (0,3) {SN3};
\node at (0,4) {$Y_{3,k}$};
\draw [ultra thick,dashed, ->] (0,0) -- (0,2.5);
\draw [ultra thick, ->] (0,3.7) -- (0,3.5);
\def\x{0.6}
\def\y{3-0.25}
\draw [fill=gray,draw=black,thick,text=white] (-2.12,2.12) circle [radius=0.5];
\node at (-2.12,2.12) {SN4};
\node at (-2.83,2.83) {$Y_{4,k}$};
\draw [ultra thick, ->] (-2.62,2.62) -- (-2.47,2.47);
\def\x{-2.12-1.2-0.6}
\def\y{2.12-0.25}
\draw [fill=white, ultra thick] (-3,0) circle [radius=0.5];
\node at (-3,0) {SN5};
\node[above] at (-1.5,0) {$d_{k+1}$};
\node at (-4,0) {$Y_{5,k}$};
\draw [ultra thick,dashed, ->] (0,0) -- (-2.5,0);
\draw [ultra thick, ->] (-3.7,0) -- (-3.5,0);
\def\x{-3.5-1.2}
\def\y{0.35}
\draw [fill=white, ultra thick] (-2.12,-2.12) circle [radius=0.5];
\node at (-2.12,-2.12) {SN6};
\node at (-2.83-0.2,-2.83+0.2) {$Y_{6,k}$};
\draw [ultra thick,dashed, ->] (0,0) -- (-2.47+0.7,-2.47+0.7);
\draw [ultra thick, ->] (-2.62,-2.62) -- (-2.47,-2.47);
\def\x{-2.12-1.2-0.6}
\def\y{-2.12-0.25}
\draw [fill=gray,draw=black,thick,text=white] (0,-3) circle [radius=0.5];
\node at (0,-3) {SN7};
\node at (0,-4) {$Y_{7,k}$};
\draw [ultra thick, ->] (0,-3.7) -- (0,-3.5);
\def\x{0-1.2-0.6}
\def\y{-3-0.25}
\draw [fill=white, ultra thick] (2.12,-2.12) circle [radius=0.5];
\node at (2.12,-2.12) {SN8};
\node at (2.83,-2.83) {$Y_{8,k}$};
\draw [ultra thick,dashed, ->] (0,0) -- (2.47-0.7,-2.47+0.7);
\draw [ultra thick, ->] (2.62,-2.62) -- (2.47,-2.47);
\def\x{2.12+0.6}
\def\y{-2.12-0.25}
\node [text=white] at (0,0) {FC};
\draw[->,decorate, decoration={snake, segment length=7mm, amplitude=1mm}] (5,-1.7) -- (4,-0.3);
\draw[->,decorate, decoration={snake, segment length=7mm, amplitude=1mm}] (4.4,-2) -- (3,-2.5);
\draw[->,decorate, decoration={snake, segment length=7mm, amplitude=1mm}] (6,-1.7) -- (3.6,2.4);
\end{tikzpicture}}
\vspace{-3mm}
\caption{{A WSN for distributed estimation, with FC quality feedback.
Each SN decides to either remain idle with cost $0$ or to collect and transmit to the FC the measurement $Y_{n,k}$ of $X_k$ with local measurement SNR $S_{M,n,k}$ and cost $c_{\mathrm{TX}}+\phi S_{M,n,k}$. The shared wireless channel results in collisions and packet losses. The FC, 
based on the measurements received, computes an MMSE estimate of $X_k$, $\hat X_k$, and broadcasts the instruction $\mathbf D_{k+1}$ based on the estimation quality 
achieved,
which is used by the SNs to adjust their sensing-transmission parameters for the next slot.}}
\label{fig:WSN}
\vspace{-5mm}
\end{figure}

\noindent In this section, we present the system model, whose parameters are listed in Table \ref{tab1}.
 Consider a WSN with one FC, depicted in Fig.~\ref{fig:WSN}, whose goal
 is to track a random process $\{X_k,k{\geq}0\}$
 following the  scalar linear Gaussian state space~model
  \begin{align}
\label{markovstate}
X_{k+1}=\sqrt{\alpha} X_k+Z_k,
\end{align}
 based on measurements collected  by $N_S$ nearby SNs.
In~(\ref{markovstate}), $k{\in}\mathbb N{\equiv}\{0,1,2,\dots\}$ is the slot index,
 $\alpha{\in}[0,1)$ is the \emph{time-correlation parameter}  and $Z_k{\sim}\mathcal N(0,\sigma_Z^2)$.
 We denote the statistical power of $X_k$ as $\sigma_X^2{=}\frac{\sigma_Z^2}{1-\alpha}$,
 and assume  $\sigma_X^2{=}1$, since any other value can be obtained by scaling.
  Each slot is divided in three phases:
 \begin{enumerate}
 \item \emph{FC instruction} $\mathbf D_k$, broadcasted by the FC (Sec.~\ref{FCinstruct});
 \item \emph{Sensing and transmission to FC}: each SN, given $\mathbf D_k$, selects its
 sensing-transmission action (Sec. \ref{ph2});
 \item \emph{Estimation at FC}: given the measurements collected, the FC estimates $X_k$ via Kalman filtering (Sec. \ref{P3}).
 \end{enumerate}
 
 \vspace{-3mm}
\subsection{Sensing and transmission to FC}
 \label{ph2}
\noindent  Each SN, at the beginning of slot $k$, given the instruction $\mathbf D_k$ broadcasted by the FC,
  selects (possibly, in a randomized fashion) 
  the sensing-transmission parameters $(A_{n,k},S_{M,n,k},B_{n,k})$, 
 where $A_{n,k}{\in}\{0,1\}$ is the \emph{activation} decision of SN $n$,
 $S_{M,n,k}{\geq}0$ is the \emph{local measurement SNR} specified below, and $B_{n,k}{\in}\{0,1,2,\dots,B\}$ is the \emph{channel index}.
 {If $A_{n,k}{=}0$, SN $n$ remains idle, hence $S_{M,n,k}{=}0$ (no measurement collected) and $B_{n,k}{=}0$ (no channel selected).
On the other hand, if $A_{n,k}{=}1$, then $B_{n,k}{\in}\{1,2,\dots,B\}$ and
 the measurement of $X_k$ by SN $n$ is given by}
\begin{align}
\label{Ynk}
Y_{n,k}=\gamma_{n,k}X_{k}+W_{A,n,k}+W_{M,n,k},
\end{align}
where $W_{A,n,k}{\sim}\mathcal N(0,1/S_A)$ is the \emph{ambient noise},
and $W_{M,n,k}{\sim}\mathcal N(0,1/S_{M,n,k})$ is the \emph{measurement noise} introduced by the sensing apparatus,
independent of each other, over time and across SNs,
  $S_{A}$ is the  \emph{local ambient SNR}, and $S_{M,n,k}$ is the
   \emph{local measurement SNR}, controlled by the $n$th SN, resulting in the sensing cost $\phi S_{M,n,k}$, where $\phi\geq 0$ is a constant.
   The transmission cost is denoted as $c_{\mathrm{TX}}$, common to all SNs,
   so that the overall sensing-transmission cost
is $c_{SN}(A_{n,k},S_{M,n,k}){=}A_{n,k}(c_{\mathrm{TX}}{+}\phi S_{M,n,k})$.
We define the \emph{normalized unitary sensing cost} $\theta{\triangleq}\frac{\phi}{c_{\mathrm{TX}}}$, and
 the \emph{sample average sensing-transmission cost for SN $n$} over a time horizon of length $T+1$ as
\begin{align}
\label{ctn}
C_n^{T}(A_{n,0}^T,S_{M,n,0}^T)=\frac{1}{T+1}\sum_{k=0}^{T}c_{SN}(A_{n,k},S_{M,n,k}).
\end{align}

The \emph{accuracy state} $\gamma_{n,k}$, taking values in the finite set $\Gamma$, models the ability of SN $n$ to accurately measure $X_k$.
We model it as a Markov chain with transition probability $\mathbb P(\gamma_{n,k+1}{=}\gamma_2|\gamma_{n,k}{=}\gamma_1){=}P_\gamma(\gamma_1;\gamma_2)$ and steady state distribution $\pi_{\gamma}(\gamma)$,
i.i.d. across SNs, and we let $\boldsymbol{\gamma}_k{=}(\gamma_{1,k},\gamma_{2,k},\dots,\gamma_{N_S,k})$.
  We denote the best accuracy state as $\gamma_{\max}{=}\max\Gamma$,
and, without loss of generality, we assume $\gamma_{\max}{=}1$ and $\pi_{\gamma}(\gamma_{\max}){>}0$.
 We denote the general scenario where $\gamma_{n,k}$ follows a Markov chain as \emph{Markov-}$\gamma$ scenario,
and the special cases where $\gamma_{n,k}{=}\gamma_{\max},\forall n,k$  deterministically 
and $\gamma_{n,k}$ is i.i.d. over time as \emph{best-}$\gamma$ and \emph{i.i.d.-}$\gamma$ scenarios, respectively.
   The $N_S$ SNs share 
 a set of $B\leq N_S$ orthogonal single-hop wireless channels to report their measurements to the FC.
 We employ the collision channel model, \emph{i.e.},
 the transmission on a given channel is successful if and only if one SN transmits in that channel. 

   
\subsection{MMSE estimator at the FC via Kalman filtering}
\label{P3}
\noindent  Let $O_{n,k}$ be the \emph{transmission outcome} for SN $n$, \emph{i.e.},
   $O_{n,k}=1$ if and only if its transmission is successful. Then, the weighted average measurement
   \begin{align}
   \label{bari}
\bar Y_{k}\triangleq\frac{\sum_{n}O_{n,k}\frac{S_{n,k}}{\gamma_{n,k}}Y_{n,k}}{\sum_{n}O_{n,k}S_{n,k}}
   \end{align}
   is a sufficient statistic for $X_k$,
   where we have defined the \emph{local SNR} for SN $n$
      \begin{align}
      \label{Slocal}
   S_{n,k}=\frac{\mathbb E[(\gamma_{n,k}X_{k})^2|\gamma_{n,k}]}{\mathbb E[(W_{A,n,k}+W_{M,n,k})^2]}=
  \gamma_{n,k}^2 \frac{S_{A}S_{M,n,k}}{S_{A}+S_{M,n,k}}.
   \end{align}   
    Given the transmission outcome and $X_k$,
   $\bar Y_k$ is a Gaussian random variable with mean $X_k$ and variance
   $\Lambda_k^{-1}$,
where we have defined the \emph{aggregate SNR} collected at the FC as
\begin{align}
\label{Stot}
\Lambda_k\triangleq\sum_{n=1}^{N_S}O_{n,k}S_{n,k}.
\end{align}

 Let $\hat X_{k-1}$ and $\hat V_{k-1}$ be the posterior mean (\emph{i.e.}, the MMSE estimate)
and variance of $X_{k-1}$ at the FC at the end of slot $k{-}1$,
\emph{i.e.}, $X_{k-1}{\sim}\mathcal N(\hat X_{k-1},\hat V_{k-1})$ is the belief of the FC
of $X_{k-1}$.
 Before collecting the measurements from the SNs in slot $k$,
 using~(\ref{markovstate}),
 the belief of the FC of $X_{k}$ is $X_{k}{\sim}\mathcal N(\sqrt{\alpha} \hat X_{k-1},V_k)$,
 where $V_k$ is the \emph{prior variance} of $X_k$,
defined recursively as
\begin{align}
\label{nu}
&V_k=\alpha \hat V_{k-1}+\sigma_Z^2=
1-\alpha(1-\hat V_{k-1})\triangleq \nu(\hat V_{k-1}).
\end{align}
Then, upon collecting the weighted average measurement $\bar Y_k$~(\ref{bari})
 with aggregate SNR $\Lambda_k$,
the FC updates the \emph{posterior variance} $\hat V_k$ and  mean $\hat X_{k}$ of $X_{k}$ as
\begin{align}
\label{nu2}
\left\{\begin{array}{l}
\hat V_k=\frac{V_k}{1+V_k\Lambda_k}
\triangleq \hat \nu(V_{k},\Lambda_k),
\\
\hat X_{k}=\sqrt{\alpha} \hat X_{k-1}+\Lambda_k\hat V_k\left(\bar Y_k -\sqrt{\alpha}\hat X_{k-1}\right).
\end{array}\right.
\end{align}
The function $\nu(\hat V_{k-1})$ determines  the prior variance of $X_k$, given the posterior variance of $X_{k-1}$,
whereas $\nu(V_k,\Lambda_k)$ determines the posterior variance of $X_k$, given its prior variance $V_k$,
as a function of the aggregate SNR $\Lambda_k$ collected at the FC.
The MSE in slot $k$ is thus
\begin{align}
&\mathbb E\left[\left.(\hat X_{k}- X_{k})^2\right|V_{k},\Lambda_k\right]
=\hat \nu(V_{k},\Lambda_k).
\end{align}
We define the \emph{sample average MSE} under $\Lambda_0^T$ over a time horizon of length $T+1$ as
\begin{align}
\label{RT}
R_T(V_0;\Lambda_0^T)=\frac{1}{T+1}\sum_{k=0}^{T}\hat V_k,
\end{align}
where $\hat V_k=\hat\nu\left(\nu(\hat V_{k-1}),\Lambda_k\right)$.

\vspace{-3mm}
\subsection{FC instruction policy}
\label{FCinstruct}
\begin{table}[t]

\caption{FC instruction  policy}
\label{tabins}
\vspace{-7mm}
\begin{center}
\footnotesize
\scalebox{0.83}{
\begin{tabular}{|c| c | c | c |}
\hline\T\B 
\multirow{2}{*}{\em Scheme} & \multirow{2}{*}{\em Activity $A_{n,k}$}  & \em Local measurement & \multirow{2}{*}{\em Channel ID $B_{n,k}$}\\
&  &\em SNR $S_{M,n,k}$ & 
\\\hline\T\B 
{\bf Coordinated}    & Centralized, $@$ FC & Centralized, $@$ FC & Centralized, $@$ FC
\\\hline\T\B 
\multirow{2}{*}{{\bf Decentralized}} & Local, w.p. $q_k(\omega_{n,k})$ & Local, $\sim S_{M,k}(\omega_{n,k})$ & \multirow{2}{*}{Local, random}
\\
 & $q_k(\cdot)$ given by FC & $S_{M,k}(\cdot)$  given by FC  &
\\
\hline
\end{tabular}
}
\end{center}
\vspace{-7mm}
\end{table}
\noindent At the beginning of each slot $k$,
the FC broadcasts an
\emph{instruction} $\mathbf D_k\in\mathcal D$, which, together with the local accuracy state $\gamma_{n,k}$,
is employed by SN $n$ to select $(A_{n,k},S_{M,n,k},B_{n,k})$. We consider the following schemes:
\subsubsection{Coordinated scheme}
  In the coordinated scheme, given $\boldsymbol{\gamma}_k$, the FC schedules the sensing-transmission action $(A_{n,k},S_{M,n,k},B_{n,k})$ of  each SN.
Note that each SN is required to report its accuracy state to the FC,
whenever its value changes.
The communication overhead required to collect such information at the FC is analyzed in Part I.
Therefore, the instruction takes the form $\mathbf D_k{=}(d_{1,k},d_{2,k},\dots,d_{N_{S},k})$, where
$d_{n,k}{=}(A_{n,k},S_{M,n,k},B_{n,k})$.
Since $\boldsymbol{\gamma}_{k}$ is perfectly known at the FC at the beginning of slot $k$,
 letting $\pi_{\boldsymbol{\gamma},k}$ be the belief of $\boldsymbol\gamma_{k}$ at the FC, we have that
   $\pi_{\boldsymbol{\gamma},k}(\boldsymbol\gamma){=}\chi(\boldsymbol\gamma{=}\boldsymbol\gamma_{k})$,
  where $\chi(\cdot)$ is the indicator function. 
The value $\mathbf D_k$ is selected 
based on $V_k$, and $\pi_{\boldsymbol{\gamma},k}$
according to some (possibly, non-stationary) \emph{instruction policy} $\delta_k(\mathbf d|V_k,\pi_{\boldsymbol{\gamma},k})\triangleq\mathbb P(\mathbf D_{k}=\mathbf d|V_k,\pi_{\boldsymbol{\gamma},k})$.
 \subsubsection{Decentralized scheme}
 In the decentralized scheme, the FC specifies 
 $\mathbf D_k{=}(q_k(\cdot),S_{M,k}(\cdot))$,
  where $q_k{:}\Gamma{\mapsto}[0,1]$ and $S_{M,k}{:}\Gamma{\mapsto}[0,\infty)$
  are, respectively,
the activation probability and 
the local measurement SNR functions  employed by each SN to select their sensing-transmission strategy in a decentralized manner, as a function of the local accuracy state $\gamma_{n,k}$.
Therefore, $\mathbf D_k$ takes value in the set $\mathcal D{\equiv}([0,1]^\Gamma{\times}\mathbb R_+^\Gamma)$,
and is generated according to some (possibly, non-stationary) policy $\delta_k(\mathbf d|V_k,\pi_{\boldsymbol\gamma,k}){\triangleq}\mathbb P(\mathbf D_{k}{=}\mathbf d|V_k,\pi_{\boldsymbol{\gamma},k})$, where
$\pi_{\boldsymbol\gamma,k}(\boldsymbol\gamma_k){=}\mathbb P(\boldsymbol\gamma_k|\mathcal H_k)$
is the belief state of the accuracy state vector $\boldsymbol\gamma_k$, given the history of observations collected up to time $k$ at the FC, $\mathcal H_k$. 
  Given 
$\mathbf D_k{=}(q_k(\cdot),S_{M,k}(\cdot))$
and the local accuracy state $\gamma_{n,k}$,
 SN $n$ chooses its action $(A_{n,k},S_{M,n,k},B_{n,k})$
as $A_{n,k}{=}1$ with probability $q_k(\gamma_{n,k})$,
$A_{n,k}{=}0$ otherwise; if $A_{n,k}{=}1$, then
$S_{M,n,k}{=}S_{M,k}(\gamma_{n,k})$ and $B_{n,k}$ is chosen uniformly from the set of  channels $\{1,2,\dots, B\}$
(if $A_{n,k}{=}0$, then $S_{M,n,k}{=}B_{n,k}{=}0$).
Due to the randomized channel accesses, this scheme may result in collisions among SNs.
The distribution of the number of successful transmissions when each SN transmits with probability $q$ is denoted as
$p_R(r;q)$, and its distribution is characterized in \cite[Prop.~4]{MichelusiP1} and, for the case $N_S\to\infty$, in \cite[Corollary 1]{MichelusiP1}.

\vspace{-3mm}
\subsection{Performance metrics and optimization problem}
\label{sec:optprob}
\noindent Given the initial prior variance and distribution  $(V_0,\pi_{\boldsymbol\gamma,0})$, and the instruction policy $\delta$,
we define the average MSE and sensing-transmission cost of SN $n$  over a finite horizon of length $T+1$
as 
\begin{align}
\label{Cest}
&\bar M_{\delta}^{T}(V_0,\pi_{\boldsymbol\gamma,0})=\mathbb E\left[\left.R_T(V_0;\Lambda_0^T)\right|
V_0,\pi_{\boldsymbol\gamma,0}\right],
\\\label{CSN}
& \bar C_{\delta}^{T,n}(V_0,\pi_{\boldsymbol\gamma,0})=\mathbb E\left[\left.
C_n^{T}(A_{n,0}^T,S_{M,n,0}^T)
\right|V_0,\pi_{\boldsymbol\gamma,0}\right],
\end{align}
where $R_T(V_0;\Lambda_0^T)$ is the sample average MSE given by (\ref{RT}),
and $C_n^{T}(A_{n,0}^T,S_{M,n,0}^T)$
is the \emph{sample average sensing-transmission cost for SN $n$}, given by (\ref{ctn}).
The expectation is computed with respect to the activation, local measurement SNR, accuracy state and medium access processes 
$\{\mathbf D_k,A_{n,k},S_{M,n,k},\gamma_{n,k},O_{n,k},n\in\{1,2,\dots,N_S\},k\in\mathbb N\}$, induced by policy $\delta$.
 In particular, we are interested in the infinite horizon $T{\to}\infty$  (average long-term performance)
 and $V_0{=}~1$,
 so that we will drop the dependence  on $T$, $V_0$ and $\pi_{\boldsymbol\gamma,0}$ in the following treatment, whenever possible.

In Part I,
we have studied the problem of determining the optimal instruction policy $\delta^*$ such that
\begin{align}
\label{optproblag}
\!\!\!\!\delta^*=&\arg\min_{\delta} \bar M_{\delta}
+\frac{\lambda}{c_{\mathrm{TX}}}\sum_{n=1}^{N_S}\bar C_{\delta}^{n},
\end{align}
where $\lambda\geq 0$ is the Lagrange multiplier, which trades off MSE and sensing-transmission cost.
The problem (\ref{optproblag}) can be solved
via DP \cite{Bertsekas2005}.
Due to the high dimensional optimization involved, in Part I
we have derived structural properties of $\delta^*$ for the
\emph{best-}$\gamma$ scenario, by exploiting the \emph{statistical symmetry} of the WSN and the \emph{large network} approximation,
based on which DP can be solved more efficiently. For the coordinated scheme, we have also
shown that a constant policy which collects a constant aggregate SNR sequence $\Lambda_k=\bar\Lambda,\forall k$ in each slot is optimal
in some special cases \cite[Theorem 2]{MichelusiP1}. We have then extended these results to the \emph{Markov-}$\gamma$ scenario.

\vspace{-5mm}
\subsection{{Complexity of DP}}
\label{complanal}
\noindent Despite the significant computational reduction achieved by exploiting
 the statistical symmetry and large network approximation, DP has high complexity.
In fact, the optimization problem in each DP stage is non-convex, and the action space is very large.
Specifically, the DP algorithm for the coordinated scheme,\footnote{We remark that, owing to the large network approximation, the DP algorithms
are defined only in the \emph{best}-$\gamma$ scenario, where the belief $\boldsymbol\gamma_{k}$ is constant, 
 based on which an heuristic scheme is defined
for the \emph{Markov}-$\gamma$ scenario, see Part I.} provided here for convenience, is given by

\noindent {\bf COORD-DP: DP algorithm for the coordinated scheme, \emph{best-}$\gamma$ scenario.}
 For $k=T,T-1,\dots,0$, solve, $\forall V_k{\in}[1-\alpha,1]$,
  \begin{align}\label{DPgenCOORD}
  \nn
&\bar W^{T-k}(V_{k})
=\!\!\!\!\min_{\Lambda_k\in [0,BS_A)}
\bar W^{T-k-1}(\nu(\hat \nu(V_{k},\Lambda_{k})))
\\&
+\hat \nu(V_{k},\Lambda_k)
+\frac{\lambda}{c_{\mathrm{TX}}} t^*(\Lambda_k)c_{SN}\left(1,S_{M}^*(\Lambda_k)\right),
\end{align}
where $\bar W^{-1}(V_{T+1})=0$, and $(t^*(\Lambda_k),S_{M}^*(\Lambda_k))$ are given in \cite[Prop.~3]{MichelusiP1}. 
The optimizer, $\Lambda_k^*(V_k)$, is the optimal aggregate SNR collected at the FC in slot $k$,
from which the optimal number of SNs activated is $t_k(V_k){=}t^*(\Lambda_k^*(V_k))$,
with local measurement SNR $S_{M,n,k}(V_k)=S_{M}^*(\Lambda_k^*(V_k))$.
\hfill\QED

 In order to implement the above DP algorithm,
the cost-to-go function $\bar W^{T-k}(V_{k})$ is evaluated only in
$N_V$ equally spaced sample points, rather than the  interval $[1{-}\alpha,1]$,
\emph{i.e.},
 \begin{align}
 \mathcal V\equiv\left\{1-\alpha+\frac{i}{N_V-1}\alpha,\ \forall i=0,1,\dots, N_V-1\right\}.
 \end{align}
For each sample point $V_k\in\mathcal V$, the optimal aggregate SNR $\Lambda_k^*(V_k)$ can be determined approximately
as follows:
 first, the space $[0,BS_A)$ is quantized into $N_L$ equally spaced  points,
 \begin{align}
 \mathcal L\equiv\left\{\frac{i}{N_L}BS_A,\ \forall i=0,1,\dots, N_L-1\right\}
 \end{align}
  (the sample point $BS_A$ is not included since it correspond to an infinite local measurement SNR, which is unfeasible). 
  Assuming an approximation of the cost-to-go function $\bar W^{T-k-1}(V_{k+1}),\ V_{k+1}{\in}\mathcal V$ in (\ref{DPgenCOORD}) 
  is available from the previous DP stages, the term $\bar W^{T-k-1}(\nu(\hat \nu(V_{k},\Lambda_{k})))$ in (\ref{DPgenCOORD}) 
  can then be approximated via linear interpolation.
An approximation of $\Lambda_k^*(V_k)$ can then be obtained via exhaustive search over the set $\mathcal L$,
with precision roughly given by $\Delta_L=BS_A/N_L$.\footnote{However, notice that, since the cost function in (\ref{DPgenCOORD}) is generally non-convex, the precision of such solution cannot be guaranteed.}
 Therefore, in order to accomplish a target precision $\Delta_L$,
  each DP stage requires $BS_AN_V/\Delta_L$ evaluations of the cost-to-go function.
 If $T_{DP}$ stages are performed, the overall complexity scales with $BS_AN_VT_{DP}/\Delta_L$.
 
 Similarly, the DP algorithm for the decentralized scheme is given by
 
 \noindent {\bf DEC-DP: DP algorithm for the decentralized scheme, \emph{best-}$\gamma$ scenario.}
 For $k=T,T-1,\dots,0$, solve, $\forall V_k{\in}[1-\alpha,1]$,
  \begin{align}
  \label{DPzeta}
&\!\!\!\!\bar W^{T-k}(V_{k})\!\!=\!\!\!\!\!\!\!\!\min_{\!\!\!\zeta\in[0,1],S_{M}}\!
\sum_{r=0}^Bp_R(r;\zeta)\hat \nu\left(\!\!V_{k},r\frac{S_AS_M}{S_A\!\!+\!\!S_M}\!\right)
\!\!+\!\!\frac{\!\lambda\zeta\!}{\!c_{\mathrm{TX}}\!} c_{SN}(1,\!S_{M}\!)
\nonumber
\\&
\!+\!\sum_{r=0}^Bp_R(r;\zeta)
\bar W^{T-k-1}\left(\nu\left(\hat \nu\left(V_{k},r\frac{S_AS_M}{S_A+S_M}\right)\right)\right),
\end{align}
where $\bar W^{-1}(V_{T+1}){=}0$, $\zeta{=}qN_S/B$
is the \emph{normalized activation probability per channel}, and
$p_R(r;\zeta)$ is the distribution of $R_k$ for $N_S{\to}\infty$ \cite[Corollary 1]{MichelusiP1}.
The optimizer, $(\zeta_k^*(V_k),S_{M,k}^*(V_k))$, is the optimal normalized activation probability and local measurement SNR in slot $k$,
from which the activation probability is given by $q_k^*(V_k){=}B\zeta_k^*(V_k)/N_S$.
\hfill\QED

In this case, for each $V_k{\in}\mathcal V$,
 an approximation of the optimal $(\zeta_k^*(V_k),S_{M,k}^*(V_k))$ can be obtained via exhaustive search over the grid 
 $[(\mathcal Z\setminus\{0\})\times\mathcal S_M]\cup\{(0,0)\}$, where 
 \begin{align}
 &\mathcal Z\equiv\left\{\frac{i}{N_Z-1},\ \forall i=0,1,\dots, N_Z-1\right\},
 \\
 &\mathcal S_M\equiv\left\{\frac{i+1}{N_M-i}S_A,\ \forall i=0,1,\dots, N_M-1\right\},
 \end{align}
and $N_Z$, $N_M$ are the number of samples.
 Note that the choice of the samples for the local measurement SNR, $\mathcal S_M$, is such that the
 interval of feasible values for the local  SNR~(\ref{Slocal}), $(0,S_A)$, is uniformly quantized.
 The points $\{0\}\times\mathcal S_M$ are not included in the search grid, since, when the transmission probability is zero, all SNs are inactive and
 their local measurement SNR is $0$.
 Similarly, $0{\notin}\mathcal S_M$, since the measurements collected with local measurement SNR $0$ are not informative and do not need to be transmitted.
 The precision in the evaluation of $\zeta_k^*(V_k)$ is roughly $\Delta_Z{=}1/(N_Z{-}1)$,
 whereas the optimal local SNR (\ref{Slocal}) is evaluated with precision roughly given by $\Delta_M{=}S_A/(N_M{+}1)$.
 Each DP stage thus involves $N_V[(N_Z-1)N_M+1]$ evaluations of the cost-to-go function (\ref{DPzeta}),
 so that the overall complexity after $T_{DP}$ stages scales approximately as $N_VT_{DP}S_A/(\Delta_Z\Delta_M)$.

Since the SNs typically have limited computational capability, in this paper, we focus on low-complexity control policies,
which can be implemented in practical systems.
Specifically, we investigate the \emph{myopic policy} (MP), defined as the solution of the optimization problem
\begin{align}
\label{MPgen}
&\delta^{(MP)}(V_k,\pi_{\boldsymbol\gamma,k})=\arg\min_{\delta}
\mathbb E\left[
\hat \nu(V_{k},\Lambda_k)
\vphantom{\sum_{n=1}^{N_S}}\right.\\&\left.\left.\nonumber
+\frac{\lambda}{c_{\mathrm{TX}}}\sum_{n=1}^{N_S}c_{SN}\left(A_{n,k},S_{M,n,k}\right)
\right|V_k,\pi_{\boldsymbol\gamma,k},\delta
\right],
\end{align}
where $\delta$ depends on the specific scheme considered,
and the expectation is 
computed with respect to the aggregate SNR collected at the FC, induced 
by policy $\delta$, and the sensing-transmission decisions of the SNs.
Such policy neglects the impact of the current decision on the future, and only optimizes the current cost,
hence it corresponds to the first DP stage ($T_{DP}{=}1$).
In particular, the overall cost balances the expected MSE in slot $k$, $\mathbb E[\hat \nu(V_{k},\Lambda_k)|V_k,\pi_{\boldsymbol\gamma,k},\delta]$,
and the expected sensing-transmission cost incurred by each SN in slot $k$, $\mathbb E\left[c_{SN}\left(A_{n,k},S_{M,n,k}\right)|V_k,\pi_{\boldsymbol\gamma,k},\delta\right]$.
We denote the average long-term MSE and sensing-transmission cost under the MP, for a specific value of $\lambda$, as $\bar M_{MP}^\lambda$ and $\bar C_{MP}^\lambda$, respectively.
\vspace{-3mm}
\begin{remark}
\label{rem1}
We note the following beneficial property of the MP:
given $V_k$ and $\Lambda_k$, the next state is $V_{k+1}{=}\nu(\hat\nu(V_k,\Lambda_k)){=}1{-}\alpha(1{-}\hat\nu(V_k,\Lambda_k))$;
therefore, the minimization of the expected MSE $\mathbb E[\left.\hat \nu(V_{k},\Lambda_k)|V_k,\pi_{\boldsymbol\gamma,k}\right|\delta]$,
implicit in the definition of the MP (\ref{MPgen}), also yields a minimization of the expected prior variance in the next slot, 
$\mathbb E[\left.\nu(\hat\nu(V_k,\Lambda_k))|V_k,\pi_{\boldsymbol\gamma,k}\right|\delta]$, \emph{i.e.}, the MP not only minimizes the present cost in slot $k$,
but, on average,  also moves  the system to a "good" next state associated to a more accurate estimate of $X_{k+1}$.
Furthermore, note that the MP is optimal when the process $X_k$ is i.i.d. ($\alpha{=}0$) and $\boldsymbol{\gamma}_k$ is i.i.d. over time.
In fact, in this case, the sensing-transmission decision in slot $k$ does not affect the next state $V_{k+1}$ and the future cost, hence $V_k=1$ in each slot.
\end{remark}
\section{Myopic Policy: Coordinated scheme}
\label{analysisCoord}
\noindent In this section, we analyze the MP for the coordinated scheme.
As in Part I, we first investigate the \emph{best-}$\gamma$ scenario, and then extend the analysis
 to the \emph{Markov-}$\gamma$ scenario.
\vspace{-2mm}
\subsection{\emph{Best-}$\gamma$ scenario}
\label{bestomegacoord}
\noindent In this case, the belief $\pi_{\boldsymbol{\gamma},k}$ is constant and  can be neglected.
From (\ref{MPgen}), using the structural properties of \cite[Prop.~2]{MichelusiP1}, \emph{i.e.}, $S_{M,n,k}=S_{M,k},\forall k$, the MP is defined as
\begin{align}
\label{MPcost}
(t^{(MP)},S_M^{(MP)})(V_k)=&
\!\!\!\!\!\!\!\!\!\!\underset{t\in\{0,1,\dots,B\},S_M\geq 0}{\arg\min}
\hat \nu\left(V_{k},t\frac{S_AS_M}{S_A+S_M}\right)
\nonumber\\&
+\frac{\lambda}{c_{\mathrm{TX}}} t c_{SN}\left(1,S_M\right),
\end{align}
where $t^{(MP)}(V_k)$ is the number of SNs activated and $S_M^{(MP)}(V_k)$
is the common local measurement SNR. 
The $t^{(MP)}(V_k)$ SNs are selected randomly from the set of $N_S$ SNs.
The following theorem derives a closed-form expression of the MP.
We denote by $\lceil x\rceil$ for $x\in\mathbb R$ the ceiling operation.
\vspace{-7mm}
\begin{thm}
\label{lemMPclosedform}
Let $\lambda{\leq}\frac{1}{\left(\sqrt{1+1/S_A}+\sqrt{\theta}\right)^2}{\triangleq}\lambda_{\mathrm{th}}$,
$v_{\mathrm{th}}(\lambda,-1){\triangleq}0$,
$t^*{\triangleq}\left\lceil\sqrt{\frac{1}{\lambda S_A}{+}\frac{1}{4}}{-}\frac{3}{2}\right\rceil$,
and, for $0\!\leq\!t\!\leq\!t^*$,
\begin{align}
\label{vth}
&v_{\mathrm{th}}(\lambda,t)
\triangleq
 \frac{
\sqrt{\lambda\theta}+\lambda\left(t+\frac{1}{2}\right)
 }
 {1-\lambda (t+1)tS_A}
\\& \nonumber
 +
  \frac{
 \sqrt{\lambda}
 \sqrt{
 \sqrt{\lambda\theta}(2t+1)
 +\lambda \theta(t+1)tS_A
  +\frac{\lambda}{4}
 +\frac{1}{S_A}
 }
 }{1-\lambda(t+1)tS_A}.
 \end{align}
 We have the following cases:
 
\noindent \emph{i)} if $V_k{>}v_{\mathrm{th}}(\lambda,t^*)$, then $t^{(MP)}(V_k){=}\min\{t^*+1,B\}$;

\noindent \emph{ii)} if $V_k{=}v_{\mathrm{th}}(\lambda,\hat t)$, for some $\hat t{\in}\{0,1,\dots,t^*\}$, then $t^{(MP)}(V_k){=}\min\{\hat t{+}1,B\}$ with probability $p_{\hat t}$,
 $t^{(MP)}(V_k){=}\min\{\hat t,B\}$ otherwise, for some $p_{\hat t}{\in}[0,1]$;

\noindent \emph{iii)} otherwise, $t^{(MP)}(V_k){=}\min\{\hat t,B\}$, where  $\hat t$ is the unique $\hat t{\in}\{0,1,\dots,t^*\}$ such that  $v_{\mathrm{th}}(\lambda,\hat t{-}1){<}V_k{<}v_{\mathrm{th}}(\lambda,\hat t)$. 

\noindent \emph{iv)} In all cases,
\begin{align}
\label{optSM}
S_M^{(MP)}(V_k)=\left(\frac{1}{\sqrt{\lambda\theta}}-\frac{1}{V_k}\right)\frac{S_AV_k}{1+t^{(MP)}(V_k)S_AV_k}.
\end{align}
\end{thm}
\noindent\emph{Proof:}
See Appendix \ref{proofoflemMPclosedform}.
\hfill\QED

\noindent Note that, when $V_k{=}v_{\mathrm{th}}(\lambda,\hat t)$, for some $\hat t{\in}\{0,1,\dots,t^*\}$, the choice of $t^{(MP)}(V_k)$ is probabilistic.
This is because both solutions $t^{(MP)}(V_k){=}\min\{\hat t,B\}$ and $t^{(MP)}(V_k){=}\min\{\hat t{+}1,B\}$
attain the same cost in (\ref{MPcost}). By varying the probability $p_{\hat t}{\in}[0,1]$, different trade-offs between MSE and sensing-transmission cost are obtained. 
The case $\lambda{>}\lambda_{\mathrm{th}}$ is of no interest, since the sensing-transmission cost
in (\ref{MPcost}) becomes too large, thus forcing the trivial MP $t^{(MP)}(V_k){=}0,\forall V_k$.

The threshold $v_{\mathrm{th}}(\lambda,t)$ is an increasing function of $t$.
The implication is that, the poorer the estimate of $X_k$, \emph{i.e.}, the larger $V_k$, the more SNs activated,
and thus the larger the sensing-transmission costs incurred. 
In other words, the limited resources available are allocated only when the FC is most uncertain about the state, \emph{i.e.}, when the estimate of
 $X_k$ is poor and needs to be improved. On the other hand, the SNs are kept idle when the FC has an accurate estimate of $X_k$, in order to preserve energy.
Moreover, $S_M^{(MP)}(V_k)$ is a piecewise increasing function of $V_k$, except at the boundaries $v_{\mathrm{th}}(\lambda,t)$ corresponding to transitions in the number of SNs activated, increasing function of $S_A$ and decreasing function of $\theta$.
In fact, $S_A$ determines the error floor in the measurement collected by each SN, so that,
as $S_A$ increases and the ambient noise becomes less relevant,
or the sensing cost decreases (as a consequence of decreasing $\theta$),
 there is a stronger incentive to collect more accurate measurements.

The next proposition gives properties of the performance achieved by the MP, in the asymptotic regime $\lambda\to \{0,\lambda_{\mathrm{th}}\}$.
\begin{propos}
\label{lemperformance}
In the limits $\lambda\to 0$ and $\lambda\to \lambda_{\mathrm{th}}$,
 the MP attains the following average long-term performance:
\begin{align}
&\lim_{\lambda\to0}\bar M_{MP}^\lambda=\hat\nu^*(BS_A),\ 
&\lim_{\lambda\to0}\bar C_{MP}^\lambda=\infty,\\
&\lim_{\lambda\to\lambda_{\mathrm{th}}}\bar M_{MP}^\lambda=1,\ 
&\lim_{\lambda\to\lambda_{\mathrm{th}}}\bar C_{MP}^\lambda=0,
\end{align}
where
\begin{align}
\label{nuing}
\!\!\hat\nu^*(x)\triangleq\frac{
\sqrt{\!(1\!-\!\alpha)^2(1\!+\!x^2)\!+\!2(1\!\!-\!\!\alpha^2)x}
\!-\!(1\!-\!\alpha)(1\!+\!x)
}
{
2\alpha x
}.
\end{align}
\end{propos}
\noindent\emph{Proof:}
See  Appendix \ref{proofoflemperformance}.
\hfill\QED

\noindent As expected, when $\lambda{\to}\lambda_{\mathrm{th}}$, the sensing-transmission cost becomes dominant in the overall MP cost function,
hence the SNs are forced to remain idle in each slot. The resulting sensing-transmission cost is zero, and the MSE is $1$, since no measurements are received at the FC.
On the other hand, when $\lambda{\to}0$, the MSE cost becomes dominant. In this case, all $B$ channels are used to transmit 
the measurements to the FC in each slot, and each measurement is collected with infinitely large measurement SNR $S_M{\to}\infty$, so that the aggregate SNR collected at the
FC is $BS_A$, hence the sensing-transmission cost converges to $\infty$ and the MSE  to $\hat\nu^*(BS_A)$
\cite[Prop.~7]{MichelusiP1}.

\subsubsection{Complexity of the MP}
\label{complmpcoord}
Note that the MP for the coordinated scheme can be determined in closed form, and therefore its complexity scales with $N_V$, the number of 
sample points in the prior variance state space $\mathcal V$.
Therefore, a significant complexity reduction is achieved with respect to DP (\ref{DPgenCOORD}),
with complexity  $BS_AN_VT_{DP}/\Delta_L$ (Sec. \ref{complanal}).

In the next section, we further specialize the analysis to the case $S_A\to\infty$, which provides further insights on the structure of the MP. In this case, the measurement $Y_{n,k}$ collected by SN $n$ is only subject to
additive Gaussian measurement noise, whereas the ambient noise is zero.
\vspace{-3mm}
\subsection{\emph{Best-}$\gamma$ scenario with $S_A\to\infty$}
\noindent We have the following corollary of Theorem~\ref{lemMPclosedform}.
\begin{corol}
\label{corolMPclosedform}
Let $\lambda\leq\lambda_{\mathrm{th}}=\frac{1}{\left(1+\sqrt{\theta}\right)^2}$
and
\begin{align}
v_{\mathrm{th}}(\lambda,0)
\triangleq
\sqrt{\lambda\theta}+\frac{\lambda}{2}
 +
 \sqrt{\lambda}
 \sqrt{
 \sqrt{\lambda\theta}
  +\frac{\lambda}{4}
 }.
\end{align}

\noindent \emph{i)} If  $V_k>v_{\mathrm{th}}(\lambda,0)$, then the MP is $t^{(MP)}(V_k)=1$ and
\begin{align}
\label{SAinf}
S_M^{(MP)}(V_k)=\frac{1}{\sqrt{\lambda\theta}}-\frac{1}{V_k}.
\end{align}

\noindent \emph{ii)} If $V_k{<}v_{\mathrm{th}}(\lambda,0)$, the MP is $t^{(MP)}(V_k){=}S_M^{(MP)}(V_k){=}0$.

\noindent \emph{iii)} Finally, if $V_k{=}v_{\mathrm{th}}(\lambda,0)$, the MP is
$t^{(MP)}(V_k){=}1$, $S_M^{(MP)}(V_k){=}\frac{1}{\sqrt{\lambda\theta}}{-}\frac{1}{V_k}$ with probaility $p_0$,
and
 $t^{(MP)}(V_k){=}0$, $S_M^{(MP)}(V_k){=}0$ with probability $1{-}p_0$, for some $p_0{\in}[0,1]$.
\end{corol}
Corollary \ref{corolMPclosedform}
 dictates that, when $S_A{\to}\infty$, only one SN may activate, \emph{i.e.}, the sensing-transmission burden
 is concentrated on a single SN, whereas all the other SNs remain idle.
In fact, the ambient noise provides an SNR floor in the quality of the measurement collected by each SN.
When $S_A$ is finite, \emph{i.e.}, the ambient noise is non-zero, it may be desirable to collect multiple measurements from multiple sensors, in order to average out the effect 
of the ambient noise, despite the fact that a large transmission cost may be incurred.
On the other hand, when $S_A$ is infinite, \emph{i.e.}, the ambient noise is zero, there is no need to average out the ambient noise, hence it is beneficial to 
collect a highly accurate measurement from one SN only, in order to minimize the transmission cost.
This result implies that one orthogonal channel ($B{=}1$) suffices in this case.
Alternatively, in order to collect the target aggregate SNR $\Lambda_k{>}0$, the FC should activate 
 $t{>}0$ SNs with local SNR $S_{M,n,k}{=}\Lambda_k/t$.
 The resulting overall network cost is
 $tc_{\mathrm{TX}}+\phi \Lambda_k$, minimized by $t=1$.

In the next theorem we characterize, in closed form, the performance of the MP when $S_A\to\infty$.
To this end, we define
 $\lambda_j^*$ to be the unique solution of $\eta_j(\lambda_j^*)=0$, where
 \begin{align}
 \label{etak}
 \eta_j(\lambda)\triangleq1-\alpha^{j}(1-\sqrt{\lambda\theta})-v_{\mathrm{th}}(\lambda,0),\ j\geq 0,\lambda\geq 0.
 \end{align}
In the statement of the theorem and in its proof, we make use of properties of $ \eta_j(\lambda)$ and $\lambda_j^*$, stated in Prop.~\ref{eta} in Appendix~\ref{proofofeta}.
\begin{thm}
\label{lemSAinf}
Let $S_A{=}\infty$, $J{\geq}1$, $\lambda{\in}(\lambda_{J-1}^*,\lambda_J^*]$, $\hat V^*{=}\sqrt{\lambda\theta}$.

\noindent \emph{i)} If $\lambda=\lambda_J^*$, then
\begin{align}
\label{Rinf}
&\!\!\!\bar M_{MP}^{\lambda,p_0}
\!=\!
1\!-\!\frac{\{1\!-\!\alpha^{J}[1\!-\!(1\!-\!\alpha)(1\!-\!p_0)]\}(1\!-\!\hat V^*)}{(J+1-p_0)(1-\alpha)},
\\
\label{Cinf}
&\!\!\!\bar C_{MP}^{\lambda,p_0}
\!=\!
\frac{1}{N_S(k+1-p_0)}
\left[c_{\mathrm{TX}}
+\phi\frac{1}{\hat V^*}(1-\hat V^*)
\right.
\\&
\left.\times\left(
  \!p_0\frac{1-\alpha^{J}}{1\!-\!\alpha^{J}(1\!-\!\hat V^*)}\!
+\!(1-p_0)\frac{1-\alpha^{J+1}}{1\!-\!\alpha^{J+1}(1\!-\!\hat V^*)}
\!\right)\right]\!.\!
\nonumber
\end{align}

\noindent \emph{ii)} Otherwise ($\lambda\in (\lambda_{J-1}^*,\lambda_{J}^*)$),
\begin{align}
\!\!\!\!\!\bar M_{MP}^{\lambda,1}
\!=&
1-\frac{(1-\alpha^{J})(1-\hat V^*)}{J(1-\alpha)},
\\
\!\!\!\!\!\bar C_{MP}^{\lambda,1}
\!=&
\frac{1}{N_S J}\left[c_{\mathrm{TX}}
+\phi\frac{1}{\hat V^*}
\frac{(1-\alpha^{J})(1-\hat V^*)
}{1-\alpha^{J}(1-\hat V^*)}\right].
\end{align}
\end{thm}
\noindent\emph{Proof:}
See  Appendix \ref{proofoflemSAinf}.
\hfill\QED

\noindent
{Consider the case $\lambda{\in}(\lambda_{J-1}^*,\lambda_{J}^*)$ (a similar argument holds for the case $\lambda{=}\lambda_J^*$).
The parameter $J$ represents the \emph{transmission period}, \emph{i.e.}, one SN is activated once every $J$ slots, whereas all SNs stay idle in the 
remaining $J{-}1$ slots.
On the other hand, $\hat V^*$ is the minimum posterior variance achieved when one SN is activated and its measurement is collected at the FC.
During the idle period, no measurements are collected, hence the posterior variance increases in each slot.
As discussed in \cite[Remark 5]{MichelusiP1}, this pattern of periodic transmissions with period $J$ can be reduced
by including a term which accounts for the \emph{outage event} $\hat V_k\geq\hat v_{\mathrm{th}}$ in the MP cost function.
Clearly, as $\lambda$ increases, the transmission period $J$ augments,
hence the SNs are activated less frequently resulting in a lower cost and poorer MSE performance.
Similarly, $\hat V^*$ increases since a smaller local measurement SNR is employed by the active SN (see (\ref{SAinf})).}
 By varying $(\lambda,p_0){\in}\mathcal L$, 
where
\begin{align*}
\label{}
&\mathcal L{\equiv}
\underset{j\geq 1}{\cup}
\left[
\left\{(\lambda,1):\lambda\in (\lambda_{j-1}^*,\lambda_{j}^*)\right\}
{\cup}
\left\{(\lambda_j^*,p_0):p_0\in[0,1]\right\}\right],
\end{align*}
we obtain different operational points $(\bar C_{MP}^{\lambda,p_0},\bar M_{MP}^{\lambda,p_0})$.
The next proposition states properties of the cost-MSE graph $(\bar C_{MP}^{\lambda,p_0},\bar M_{MP}^{\lambda,p_0})_{(\lambda,p_0)\in\mathcal L}$.
To this end, we define the following ordering of the elements in $\mathcal L$: let 
$(\lambda^{(i)},p_0^{(i)})\in\mathcal L$, $i=1,2$ with $(\lambda^{(1)},p_0^{(1)})\neq (\lambda^{(2)},p_0^{(2)})$;
then, $(\lambda^{(1)},p_0^{(1)})\succ (\lambda^{(2)},p^{(2)})$ if either
$\lambda^{(1)}>\lambda^{(2)}$, or $\lambda^{(1)}=\lambda^{(2)}$ and
 $p_0^{(1)}<p_0^{(2)}$.
\begin{propos}
\label{lemconvex}

\noindent \emph{i)} $(\bar C_{MP}^{\lambda,p_0},\bar M_{MP}^{\lambda,p_0})_{(\lambda,p_0)\in\mathcal L}$ is continuous.

\noindent \emph{ii)}  $\bar C_{MP}^{\lambda,p_0}$  is decreasing in $(\lambda,p_0)\in\mathcal L$,
 whereas $\bar M_{MP}^{\lambda,p_0}$  is increasing in $(\lambda,p_0)\in\mathcal L$,
 \emph{i.e.},
\begin{align}
\label{decreasing}
&\bar C_{MP}^{\lambda^{(1)},p_0^{(1)}}
<\bar C_{MP}^{\lambda^{(2)},p_0^{(2)}},\ 
\bar M_{MP}^{\lambda^{(1)},p_0^{(1)}}
>\bar M_{MP}^{\lambda^{(2)},p_0^{(2)}},
\\\nonumber&
\forall(\lambda^{(i)},p_0^{(i)})\in\mathcal L,\ i=1,2\text{ s.t. } (\lambda^{(1)},p_0^{(1)})\succ (\lambda^{(2)},p_0^{(2)}).
\end{align}
\end{propos}
\noindent\emph{Proof:}
See Appendix \ref{proofoflemconvex}.
\hfill\QED

\noindent 
Prop. \ref{lemconvex} shows a desirable property of the MP for the special case $S_A{\to}\infty$.
In particular, the larger $\lambda$, \emph{i.e.} the more resource constrained the system, the smaller the sensing-transmission cost and the larger the
MSE. The implication is that we can tune $\lambda$ in order to achieve the desired trade-off between
cost and MSE. Note that 
 (\ref{decreasing}) is not expected. In fact, the MP is designed to minimize only the instantaneous cost (\ref{MPcost}),
not the average long-term performance. 
  The more general case $S_A<\infty$ is difficult to analyze, due to the complex structure of the MP and the resulting evolution of $\{V_k,\ k\geq 0\}$.
  In the next section, we analyze the \emph{Markov-}$\gamma$ scenario.
  \vspace{-3mm}
  \subsection{\emph{Markov-}$\gamma$ scenario}
\noindent In this case, the accuracy state of each SN fluctuates over time according to a Markov chain, thus causing random fluctuations
in the aggregate SNR collected at the FC.
 The optimal policy is difficult to characterize,
 due to the high dimensionality of the problem.
 Herein, as in Part I, we define a \emph{sub-optimal coordinated MP}, based 
 on the MP derived in Sec. \ref{bestomegacoord}.
 Specifically, 
 let $r(\cdot;\boldsymbol{\gamma}_k){:}\{1,2,\dots,N_S\}{\mapsto}\{1,2,\dots,N_S\}$
be a ranking of SNs indexed by $\boldsymbol{\gamma}_k$, such that $r(m;\boldsymbol{\gamma}_k)$
is the label of the SN with the $m$th highest accuracy state,
\emph{i.e.}, $\gamma_{r(1;\boldsymbol{\gamma}_k),k}\geq\gamma_{r(2;\boldsymbol{\gamma}_k),k}\geq,\dots,\geq\gamma_{r(N_S;\boldsymbol{\gamma}_k),k}$.
 
 Let $\{\tilde V_k,k\geq 0\}$ be a \emph{virtual prior variance process}, generated as if all measurements were collected with
 the best accuracy state $\gamma_{\max}$. Starting from $\tilde V_0=V_0$, we thus have $\tilde V_{k+1}=\nu(\hat\nu(\tilde V_k,\tilde\Lambda_k))$,
 where $\tilde\Lambda_k=t^{(MP)}(\tilde V_k)\frac{S_AS_M^{(MP)}(\tilde V_k)}{S_A+S_M^{(MP)}(V_k)}$.
 We define the sub-optimal coordinated MP (SCMP) as follows.
 
  \noindent\textbf{SCMP}:
Given $\lambda{\leq}\lambda_{\mathrm{th}}$, the virtual prior variance state $\tilde V_k$, and $\boldsymbol\gamma_k$, 
the
$t^{(MP)}(\tilde V_k)$  SNs with the best accuracy state are activated  in slot $k$,
 with local measurement SNR $S_M^{(MP)}(\tilde V_k)$,
\begin{align}
\!\!\nonumber
\left\{
\begin{array}{l}
\!\!\!\!A_{r(m;\boldsymbol{\gamma}_k),k}\!=\!1,\ \!\!S_{M,r(m;\boldsymbol{\gamma}_k),k}\!=\!S_M^{(MP)}(\tilde V_k),\forall m\leq t^{(MP)}(\tilde V_k),
\\
\!\!\!\!A_{r(m;\boldsymbol{\gamma}_k),k}\!=\!0,\ \forall m>t^{(MP)}(\tilde V_k).\hfill\QED
\end{array}
\right.
\end{align}

In the \emph{best-}$\gamma$ scenario, SCMP simplifies to the MP given by Theorem~\ref{lemMPclosedform}.
 In the next proposition, we derive a bound to the average long-term performance of SCMP 
in the \emph{Markov-}$\gamma$ scenario,
$(\bar C_{MP}^{\lambda},\bar M_{MP,\lambda})$, with respect to the performance achieved
in the \emph{best-}$\gamma$ scenario, $(\bar C_{MP}^{\lambda,\gamma_{\max}},\bar M_{MP}^{\lambda,\gamma_{\max}})$.
Its proof is similar to the proof of \cite[Theorem 3]{MichelusiP1}, and is thus omitted.
\begin{propos}
Under the SCMP, if $\pi_{\gamma}(\gamma_{\max}){<}1$ 
 and $N_S{\geq}\frac{B-1}{\pi_{\gamma}(\gamma_{\max})}$, then $\bar C_{MP}^{\lambda,\gamma_{\max}}{=}\bar C_{MP}^{\lambda}$ and
\begin{align}
\label{ineq2}
\!\!\!0
\!\leq\!
\bar M_{MP}^{\lambda}\!-\!\bar M_{MP}^{\lambda,\gamma_{\max}}
\!\leq\!
\frac{\exp\left\{-\frac{\left(N_S\pi_{\gamma}(\gamma_{\max})-B+1\right)^2}{2N_S\pi_{\gamma}(\gamma_{\max})}\right\}
}{1-\alpha}.
\end{align}
\end{propos}
Note that SCMP achieves the same average long-term cost as if all the SNs could sense with the best accuracy state $\gamma_{\max}$.
This is a consequence of the fact that SCMP is generated according to the virtual prior variance  state $\tilde V_k$, whose evolution
emulates that of the \emph{best-}$\gamma$ scenario.
In the next section, we analyze the MP for the decentralized scheme.
\vspace{-3mm}
\section{Myopic Policy: Decentralized scheme}
\label{analysisDistr}
\noindent We first investigate the \emph{best-}$\gamma$ scenario, and then
 extend our analysis to the \emph{Markov-}$\gamma$ scenario.
 \vspace{-3mm}
\subsection{\emph{Best-}$\gamma$ scenario}
\label{bestomegadistr}
\noindent In the decentralized scheme, the MP is defined as
\begin{align}
\label{DPMP}
(q^{(MP)},S_{M}^{(MP)})(V_k)
=&\arg\!\!\!\!\!\!\!\!\!\!\!\min_{q\in[0,1],S_{M}\geq 0}\!\!\!\!\!
\mathbb E\left[\hat \nu\left(V_{k},\frac{R_kS_AS_{M}}{S_A+S_{M}}\right)\right]
\nonumber\\&
+\lambda N_S q (1+\theta S_M),
\end{align}
where $R_k$ is the number of packets successfully received at the FC, as a result of having each node transmit
with probability $q$ in one of the $B$ orthogonal channels available.

We focus on the \emph{large network} approximation, \emph{i.e.},
on the asymptotic  scenario of large number of SNs
$N_S\to\infty$, where we fix the \emph{normalized activation probability} $\zeta{=}qN_S/B$, and optimize over the values of $\zeta$ and $S_M$.
Then, the MP for $N_S{\to}\infty$ is defined as
\begin{align}
\label{DPMP2}
\!\!\!\!(\zeta^{(MP)},S_{M}^{(MP)})(V_k)
=\arg\!\!\!\!\!\min_{\zeta\geq 0,S_{M}\geq 0}
f(\zeta,S_M,V_k),
\end{align}
where, letting $N_S\to\infty$ in (\ref{DPMP}), we have defined
\begin{align*}
\label{}
&\!\!f(\zeta\!,\!S_M\!,\!V_k)\!\!=\!\!\!
\sum_{r=0}^B\!\mathcal B_{B}\!\left(r;\rho(\zeta)\right)\!
\hat \nu\!\!\left(\!\!V_{k},\!\!\frac{rS_AS_{M}}{S_A\!+\!S_{M}\!}\!\right)
\!\!+\!\!\lambda \zeta B(1+\theta S_M),
\end{align*}
  we have used the fact that
 $R_k$ converges to a binomial random variable with $B$ trials and success probability
$\rho(\zeta)=\zeta e^{-\zeta}$  \cite[Corollary~1]{MichelusiP1},
and we have defined the PMF of the binomial distribution $\mathcal B_{B}\left(r;\rho\right)=
\left(\begin{array}{c}B\\r
\end{array}
\right)\rho^r\left(1-\rho\right)^{B-r}
$.
The following theorem characterizes the solution of (\ref{DPMP2}).
\vspace{-3mm}
\begin{thm}
\label{lemMPdistr}
Let
$\lambda<\lambda_{\mathrm{th}}$,
where $\lambda_{\mathrm{th}}$ is defined in Theorem~\ref{lemMPclosedform},
and $v_{\mathrm{th}}(\lambda,0)$ be given by (\ref{vth}) for $t{=}0$.

\noindent\emph{i)} If $V_k\leq v_{\mathrm{th}}(\lambda,0)$,
then $(\zeta^{(MP)}(V_k),S_{M}^{(MP)}(V_k))=(0,0)$.

\noindent\emph{ii)} Otherwise, $(\zeta^{(MP)}(V_k),S_{M}^{(MP)}(V_k))=(\zeta,S_M)$
must simultaneously solve,
for some $\zeta\in (0,1)$, $S_M>0$,
\begin{align}
\label{}
\!\!\left\{
\begin{array}{l}
\!\!\!\!h(S_M\!,\!\zeta\!,\!V_k)\!\triangleq\!
-\mathbb E\!\!\left[\!
\left.\hat \nu\!\left(\!V_{k},\!\!\frac{R_kS_AS_{M}}{S_A+S_{M}}\right)^2\!\!\!\!\frac{R_kS_A^2}{(S_A+S_{M})^2}
\right|\!\rho(\zeta)\!
\right]
\!\!+\!\!\lambda \zeta B\theta\!=\!0,
\\
\!\!\!\!g(S_M,\zeta,V_k)\!\triangleq\!
\mathbb E\left[\left.\hat \nu\left(\!V_{k},\!\!\frac{R_kS_AS_M}{S_A\!+\!S_M}\right)\frac{R_k-\rho(\zeta)B}{\rho(\zeta)(1-\rho(\zeta))}\right|\!\rho(\zeta)\!
\right]
\nonumber\\
\qquad\qquad\qquad\qquad
+\lambda B\frac{e^{\zeta}}{1-\zeta} (1+\theta S_M)=0,
\end{array}
\right.
\end{align}
where the expectation is computed with respect to the PMF of $R_k\sim\mathcal B_{B}\left(\cdot;\rho(\zeta)\right)$.
Moreover,
\begin{align}
\label{boundz}
&\!\!\!0<\zeta^{(MP)}(V_k)<\min\left\{1,2\ln\left(\frac{V_k}{\sqrt{\lambda\theta}}\right)\right\}\triangleq \zeta_{\mathrm{th}}^{\max}(V_k)
\\&
\label{boundSM}
\text{and }S_{M,\mathrm{th}}^{\min}\leq S_{M}^{(MP)}(V_k)\leq S_{M,\mathrm{th}}^{\max},\ \text{where}
\end{align}
\begin{align}
&S_{M,\mathrm{th}}^{\min}\triangleq\frac{
-\lambda\theta S_A-\lambda(1+V_kS_A)+V_k^2S_A
}
{2\lambda\theta(1+V_kS_A)}
\\&
-
\frac{
\sqrt{
[(\lambda\theta+V_k^2)S_A-\lambda(1+V_kS_A)]^2
-4\lambda\theta V_k^2S_A^2
}
}
{2\lambda\theta(1+V_kS_A)},
\nonumber\\&
S_{M,\mathrm{th}}^{\max}\triangleq
\min\left\{\frac{
\!-\!\lambda\theta S_A\!-\!\lambda(1\!+\!V_kS_A)\!+\!V_k^2S_A
}
{2\lambda\theta (1+V_kS_A)}\right.
\label{smmax}
\\&
\left.\!\!+
\frac{
\sqrt{
[(\lambda\theta\!+\!V_k^2)\!-\!\lambda(1/S_A\!+\!V_k)]^2\!-\!4\lambda\theta V_k^2
}
}
{2\lambda\theta (1/S_A+V_k)},S_A\!\!\left(\!\!\frac{V_k}{\sqrt{\lambda\theta}}\!-\!1\!\!\right)\!\!\right\}.
\nonumber
\end{align}
\end{thm}
\noindent\emph{Proof:}
See Appendix \ref{proofoflemMPdistr}.
\hfill\QED

\noindent
The MP dictates that the SNs activate only when the estimation quality at the FC is poor, \emph{i.e.},
 $V_k{>}v_{\mathrm{th}}(\lambda,0)$, in order to improve the estimate, and remain idle to preserve energy when it is accurate ($V_k\leq v_{\mathrm{th}}(\lambda,0)$).
 Therefore, the MP induces an efficient utilization of the scarce resources available in the system.
 Interestingly, the threshold on the prior variance state, $v_{\mathrm{th}}(\lambda,0)$, and on the Lagrange multiplier, $\lambda_{\mathrm{th}}$,
 have the same expression as in the coordinated scheme (see Theorem~\ref{lemMPclosedform}).
 These thresholds are independent of the number of channels $B$. This is because, when $\lambda{\to}\lambda_{\mathrm{th}}$,
 the sensing-transmission cost dominates the cost function defining the MP, hence the SNs activate with (normalized) probability close to
 zero. It follows that, with high probability, only one channel will be occupied, and the remaining channels remain unused. 
 The practical implication is that, when $\lambda\to\lambda_{\mathrm{th}}$,
 \emph{i.e.}, the WSN is severely energy constrained, $B{=}1$ suffices. 
 
 Note that the MP, when $V_k{>}v_{\mathrm{th}}(\lambda,0)$, must simultaneously solve
  $h(S_{M}^{(MP)}(V_k),\zeta^{(MP)}(V_k),V_k){=}0$ and 
 $g(S_{M}^{(MP)}(V_k),\zeta^{(MP)}(V_k),V_k){=}0$. This is a set of \emph{necessary} conditions, but they may not be sufficient.
 In fact, the cost function defining the MP in (\ref{DPMP2}) is, in general, non-convex with respect to $(\zeta,S_M)$.
 We now present an iterative algorithm to determine a \emph{local} minimum of (\ref{DPMP2}), for the case $V_k>v_{\mathrm{th}}(\lambda,0)$.
 \vspace{-3mm}
 \begin{algo}\label{algoMP}$ $\\
\noindent 1) Let $S_{M}^{(0)}\in (S_{M,\mathrm{th}}^{\min},S_{M,\mathrm{th}}^{\max}),\ \zeta^{(0)}\in (0,\zeta_{\mathrm{th}}^{\max}(V_k))$, $i=0$;

\noindent 2) given $\zeta^{(i)}$, determine
\begin{align}
S_{M}^{(i+1)}=\underset{S_M\in (S_{M,\mathrm{th}}^{\min},S_{M,\mathrm{th}}^{\max})}{\arg\min}f(\zeta^{(i)},S_M,V_k)
\end{align}
as follows: if $h(S_{M,\mathrm{th}}^{\min},\zeta^{(i)},V_k){\geq}0$, set $S_{M}^{(i+1)}{=}S_{M,\mathrm{th}}^{\min}$;
if $h(S_{M,\mathrm{th}}^{\max},\zeta^{(i)},V_k){\leq}0$, set $S_{M}^{(i+1)}{=}S_{M,\mathrm{th}}^{\max}$;
otherwise, determine $S_{M}^{(i+1)}$ as the unique 
 $S_M{\in}(S_{M,\mathrm{th}}^{\min},S_{M,\mathrm{th}}^{\max})$ such that $h(S_M,\zeta^{(i)},V_k){=}0$, using the \emph{bisection method} \cite{bisection};
 
\noindent 3) given $S_{M}^{(i+1)}$, determine
\begin{align}
\zeta^{(i+1)}=\underset{\zeta\in (0,\zeta_{\mathrm{th}}^{\max}(V_k))}{\arg\min}f(\zeta,S_{M}^{(i+1)},V_k)
\end{align}
as follows: if $g(S_{M}^{(i+1)},\zeta_{\mathrm{th}}^{\max}(V_k),V_k){\leq}0$, set $\zeta^{(i+1)}{=}\zeta_{\mathrm{th}}^{\max}(V_k)$;
otherwise, determine $\zeta^{(i+1)}$ as the unique 
 $\zeta{\in}(0,\zeta_{\mathrm{th}}^{\max}(V_k))$ such that $g(S_{M}^{(i+1)},\zeta,V_k){=}0$, using the \emph{bisection method};

\noindent 4) update $i{:}i+1$ and repeat from steps 2) and 3) until convergence; return $\zeta^{(MP)}(V_k)=\zeta^{(i)}$,
$S_{M}^{(MP)}(V_k)=S_{M}^{(i)}$.
\end{algo}
Note that Algorithm \ref{algoMP} is guaranteed to converge to a \emph{local} minimum of the MP cost function (\ref{DPMP2}),
since, at each step 2-3), the function $f(\cdot)$ is minimized while keeping the other parameter fixed, and
 the MP solution $(\zeta^{(MP)}(V_k),S_{M}^{(MP)}(V_k))$ lies in the bounded set $(0,\zeta_{\mathrm{th}}^{\max}(V_k))\times(S_{M,\mathrm{th}}^{\min},S_{M,\mathrm{th}}^{\max})$.
In steps 2{-}3), we have used the fact that $h(\cdot)$ and $g(\cdot)$
are the derivatives of $f(\cdot)$  with respect to $S_M$ and $\zeta$,
and these functions are increasing in $S_M$ and $\zeta$, respectively (see Appendix~\ref{proofoflemMPdistr}).

A corollary of Theorem \ref{lemMPdistr} is given below, for the case $B{=}1$.
\begin{corol}
\label{sdfsdf}
Let $B{=}1$.

\noindent \emph{i)} If $V_k{\leq}v_{\mathrm{th}}(\lambda,0)$,
then $(\zeta^{(MP)}(V_k),S_{M}^{(MP)}(V_k))=(0,0)$.

\noindent \emph{ii)} 
Otherwise, 
\begin{align}
\label{SMMP}
S_{M}^{(MP)}(V_k)=\left(\frac{e^{-\zeta^{(MP)}(V_k)/2}}{\sqrt{\lambda\theta}}-\frac{1}{V_k}\right)\frac{V_kS_A}{1+V_kS_A},
\end{align}
and $\zeta^{(MP)}(V_k)$ is the unique $\zeta\in (0,\zeta_{\mathrm{th}}^{\max}(V_k))$ solving
\begin{align}
\label{solvez}
\!\!\!\frac{-V_kS_A}{1\!+\!V_kS_A}\!\!\left(
\!\!V_k
\!-\!e^{\frac{\zeta}{2}}\sqrt{\lambda\theta}\frac{2\!-\!\zeta}{1\!-\!\zeta}
\!+\!\frac{e^{\zeta}}{1\!-\!\zeta}\frac{\lambda\theta}{V_k}
\right)
\!+\!\frac{\lambda e^{\zeta}}{1\!-\!\zeta}
\!=\!0.\!\!
\end{align}
\end{corol}
For this case, a stronger result can be proved: the solution is a global minimum of (\ref{DPMP2}), rather than a local one for the general case $B{\geq}2$.
 $\zeta^{(MP)}(V_k)\in (0,\zeta_{\mathrm{th}}^{\max}(V_k))$ can be determined using the bisection method \cite{bisection},
 by exploiting the fact that (\ref{solvez}) is an increasing function of $\zeta$.
 Note that, for fixed $\zeta^{(MP)}(V_k)$, $S_{M}^{(MP)}(V_k)$ is an increasing function of $S_A$ and $V_k$, and decreasing function of 
 $\lambda$ and $\theta$ (however, $\zeta^{(MP)}(V_k)$ is also a function of these parameters via (\ref{solvez})). In fact, the larger $S_A$ (\emph{i.e.}, the smaller the
 error floor induced by the ambient noise) or
 $V_k$ (\emph{i.e.}, the poorer the quality of the estimate),
 or the smaller $\theta$ (\emph{i.e.}, the smaller the sensing cost) or $\lambda$ (\emph{i.e.}, the milder the cost constraint),
 the stronger the incentive to sense with higher local measurement SNR.
  By further specializing Corollary \ref{sdfsdf} to $\theta{=}0$ (no transmission cost), $S_A{=}\infty$ (no ambient noise) and $V_k{=}1{-}\alpha^{J_k+1}$, we obtain the
 MP \cite[Sec. II.B]{MichelusiP1}.
 
\subsubsection{Complexity of the MP}
\label{complmpdec}
Unlike the coordinated scheme,
the MP for the decentralized one cannot be determined in closed form.
For each $V_k\in\mathcal V$,
in order to determine $S_{M}^{(i+1)}$ in step 2) of Algorithm \ref{algoMP} with precision $\Delta_M$\footnote{The precision is evaluated with respect to the local SNR (\ref{Slocal}), in order to have a fair comparison with the analysis in Sec. \ref{complanal}} using the bisection method \cite{bisection},
at most $I_2\triangleq K_2-\log_2\Delta_M$
evaluations of $f(\zeta^{(i)},S_M,V_k)$ are needed (each corresponding to an iteration of the bisection method), where 
$K_2$ is a constant which depends on the initial search interval $[S_{M,\mathrm{th}}^{\min},S_{M,\mathrm{th}}^{\max}]$.
 Similarly, in order to determine $\zeta^{(i+1)}$ in step 3) of Algorithm~\ref{algoMP} with precision $\Delta_Z$ using the bisection method,
at most $I_3\triangleq K_3-\log_2\Delta_Z$
evaluations of $f(\zeta,S_{M}^{(i+1)},V_k)$ are needed (each corresponding to an iteration of the bisection method), where 
$K_3$ is a constant which depends on the initial search interval $[0,\zeta_{\mathrm{th}}^{\max}(V_k)]$.
For $\Delta_M,\Delta_Z\ll 1$ we thus obtain $I_2{\simeq}-\log_2\Delta_M$ 
and $I_2{\simeq}-\log_2\Delta_Z$. 
 Assuming steps 2) and 3) of Algorithm~\ref{algoMP}  are repeated $T_{MP}$ times, the
overall complexity thus scales as $-N_VT_{MP}\log_2(\Delta_M\Delta_Z)$.
We conclude that the complexity of the MP algorithm scales with the logarithm of $1/(\Delta_M\Delta_Z)$,
and thus provides a significant complexity reduction with respect to DP (\ref{DPzeta}),
whose complexity scales linearly with $1/(\Delta_M\Delta_Z)$ (Sec.~\ref{complanal}).
We have verified numerically that Algorithm \ref{algoMP} typically converges in few iterations ($T_{MP}{\sim}5$).
In the special case $B{=}1$ studied in Corollary~\ref{sdfsdf}, $S_{M}^{(MP)}(V_k)$ can be determined exactly as a function of
$\zeta^{(MP)}(V_k)$, whereas $\zeta^{(MP)}(V_k)$
 can be determined via one run of the bisection method \cite{bisection} to solve (\ref{solvez}), resulting in the overall complexity 
 $-N_V\log_2(\Delta_Z)$.

\vspace{-3mm}
\subsection{\emph{Markov-}$\gamma$ scenario}
\noindent We now discuss the \emph{Markov-}$\gamma$ scenario.
 As for the coordinated scheme, we define a \emph{sub-optimal decentralized MP} (SDMP), based 
 on the MP derived in Sec. \ref{bestomegadistr}.
 
   \noindent\textbf{SDMP}: 
Given $\lambda\leq \lambda_{\mathrm{th}}$ and
the value of $V_k$ fed back from the FC,
the activation probability is defined as
\begin{align*}
\!q^{(MP)}(V_k,\gamma)\!=\!
\left\{
\!\!\!\begin{array}{ll}
1, & \gamma>\gamma_{\mathrm{th}},\\
\frac{\frac{B}{N_S}\zeta^{(MP)}(V_k)-\sum_{\gamma>\gamma_{\mathrm{th}}}\pi_{\gamma}(\gamma)}{\pi_{\gamma}(\gamma_{\mathrm{th}})},
& \gamma=\gamma_{\mathrm{th}},\\
0, &\gamma<\gamma_{\mathrm{th}},
\end{array}
\right.
\end{align*}
and the local measurement SNR as $S_{M,n,k}{=}S_{M}^{(MP)}\!(V_k)$,
where $\gamma_{\mathrm{th}}$ uniquely solves~$\!\!\!\underset{\gamma\geq \gamma_{\mathrm{th}}}\sum\!\!\!\!\pi_{\gamma}(\gamma){\geq}
\frac{B}{N_S}\zeta^{(MP)}(V_k){>}\!\!\!\underset{\gamma>\gamma_{\mathrm{th}}}\sum\!\!\!\pi_{\gamma}(\gamma)$.
\hfill\QED

\begin{figure}[t]
\centering
\includegraphics[width = .85\linewidth,trim = 10mm 4mm 10mm 9mm,clip=true]{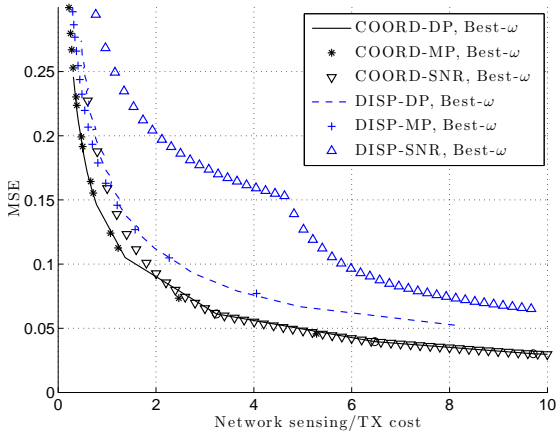}
\vspace{-3mm}
\caption{MSE as a function of the network cost, \emph{best-}$\gamma$ scenario, $N_S=20$.}
\vspace{-5mm}
\label{NS20}
\end{figure}

\noindent Note that
$\sum_{\gamma}q^{(MP)}(V_k,\gamma)\pi_{\gamma}(\gamma)N_S/B{=}\zeta^{(MP)}(V_k)$,
\emph{i.e.}, all SNs activate with \emph{marginal} normalized probability $\zeta^{(MP)}(V_k)$,
with respect to the steady state distribution of $\gamma_{n,k}$.

The performance of the sub-optimal decentralized MP is difficult to characterize. In fact, due to the Markov property of the accuracy state $\gamma_{n,k}$,
the number of collisions and successful transmissions are correlated over time.
However, the following proposition holds in the \emph{i.i.d.-}$\gamma$ scenario.
To this end, we denote by
$(\bar C_{MP}^{\lambda},\bar M_{MP}^{\lambda})$
and $(\bar C_{MP}^{\lambda,\gamma_{\max}},\bar M_{MP}^{\lambda,\gamma_{\max}})$
the performance in the
\emph{i.i.d.-}$\gamma$  and \emph{best-}$\gamma$ scenarios, respectively.
\begin{propos}
In the \emph{i.i.d.-}$\gamma$ scenario, if
$N_S\geq B/\pi_{\gamma}(\gamma_{\max})$, then 
$\bar C_{MP}^{\lambda}=\bar C_{MP}^{\lambda,\gamma_{\max}}$, $\bar M_{MP}^{\lambda}=\bar M_{MP}^{\lambda,\gamma_{\max}}$.
\end{propos}
As shown in Part I, this is a consequence of the fact that, if the conditions of the proposition hold, 
then $\gamma_{\mathrm{th}}{=}\gamma_{\max}$,
hence only the SNs with the best accuracy state may activate under SDMP, so that there is no degradation in the aggregate SNR collected at the FC, $\Lambda_k$,
compared to the \emph{best-}$\gamma$ scenario. In other words, a densely deployed WSN provides \emph{sensing diversity}.

\section{Numerical Results}\label{numres}
\noindent In this section, we provide numerical results.
Unless otherwise stated, we consider a WSN of size $N_S\in\{20,100\}$ SNs (\emph{small} and \emph{large} WSN, respectively).
We let $c_{\mathrm{TX}}{=}1$, $S_A{=}20$, $\phi{=}0.25$, $\alpha{=}0.96$, and $B{=}5$.
We consider the \emph{best-}$\gamma$ scenario only. Similar considerations hold for the \emph{Markov-}$\gamma$ scenario.
The interested reader is referred to Part~I for a numerical evaluation of the \emph{Markov-}$\gamma$ scenario.
We consider the following schemes, evaluated via Monte-Carlo simulation over $T=10^5$ slots:

\noindent$\bullet$ \emph{COORD-DP}: optimal coordinated scheme, obtained via $T_{DP}=100$ DP iterations (see Part I);
 
 \noindent$\bullet$  \emph{DEC-DP}:  optimal decentralized scheme, obtained via $T_{DP}=100$ DP iterations (see Part I);

\noindent
$\bullet$ \emph{COORD-SNR}: max coordinated aggregate SNR scheme; non-adaptive policy which maximizes the expected aggregate SNR at the FC, under cost constraints for the SNs (see Part~I);

\noindent
$\bullet$ \emph{DEC-SNR}: max decentralized aggregate SNR scheme; non-adaptive policy which maximizes the expected aggregate SNR at the FC, under cost constraints for the SNs (see Part~I);

\noindent$\bullet$ \emph{COORD-MP}: MP for the coordinated scheme (Sec. \ref{analysisCoord});
 
 \noindent$\bullet$  \emph{DEC-MP}: MP for the decentralized scheme (Sec. \ref{analysisDistr}), derived via Algorithm \ref{algoMP}.

\begin{figure}[t]
\centering
\includegraphics[width = .85\linewidth,trim = 10mm 4mm 10mm 9mm,clip=true]{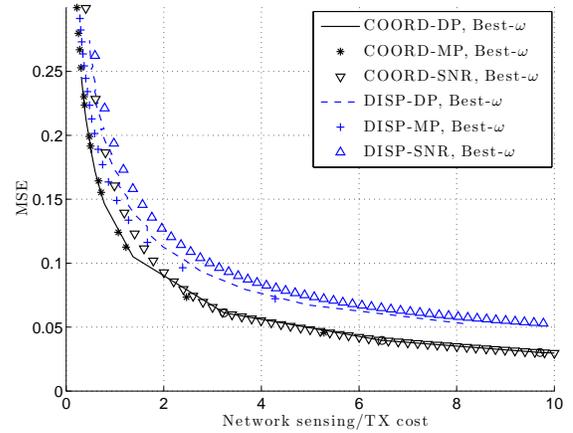}
\vspace{-3mm}
\caption{MSE as a function of the network cost, \emph{best-}$\gamma$ scenario, $N_S=100$.}
\vspace{-5mm}
\label{NS100}
\end{figure}

In Figs. \ref{NS20} and  \ref{NS100}, we plot the
 MSE (\ref{Cest}) as a function of the network sensing-transmission cost  (\ref{CSN})
 for the small and large WSN scenarios, respectively,  obtained by varying the Lagrange multiplier $\lambda$.
We notice that, in both cases, COORD-MP and DEC-MP incur no performance degradation with respect to
their DP counterparts COORD-DP and DEC-DP, respectively, at a fraction of the complexity.
As conjectured in Remark \ref{rem1}, this is because the MP not only minimizes the present cost in slot $k$,
but, on average,  also moves  the system to a "good" next state.
{Therefore, as shown in Part I, similar to the DP policies, also the MP outperforms the technique proposed in \cite{Msechu}.}
On the other hand, the non-adaptive schemes  COORD-SNR and DEC-SNR incur a significant performance degradation,
since they greedily maximize the expected aggregate SNR collected at the FC, $\mathbb E[\Lambda_k]$, but do not take into account 
the fluctuations in $\Lambda_k$, and hence, 
 in the
quality state $V_k$, resulting from cross-layer factors such as the decentralized access decisions of the SNs and the uncertain channel outcomes.

\begin{figure}[t]
\centering
\includegraphics[width = .85\linewidth,trim = 10mm 4mm 10mm 9mm,clip=true]{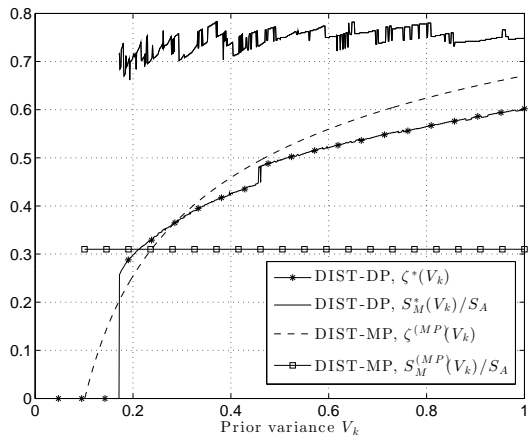}
\vspace{-3mm}
\caption{Structure of DEC-DP and  DEC-MP  as a function of the prior variance $V_k$.
The corresponding simulated network cost is $\simeq 1.66$ and the MSE is $\simeq 0.12$ for both schemes.}
\vspace{-5mm}
\label{figstruct}
\end{figure}

In Fig. \ref{figstruct}, we plot the structure of DEC-DP and DEC-MP as a function of the quality state $V_k$.
We note that, as $V_k$ increases,
\emph{i.e.}, the estimate of $X_k$ is less accurate,
 both $\zeta^*(V_k)$ and $\zeta^{(MP)}(V_k)$ increase,
 in order to achieve a higher estimation accuracy.
 On the other hand, when the estimation accuracy is good ($V_k<0.2$ for DEC-DP and $V_k<0.1$ for DEC-MP),
  the activation probability is zero, so that the SNs can
 save energy.
 The threshold on the estimation quality below which the SNs remain idle, $v_{\mathrm{th}}(\lambda,0)$, is given in closed form by (\ref{vth}) for $t{=}0$.
Note that the normalized activation probability is larger for DEC-MP than for DEC-DP. The resulting higher transmission cost for the former is balanced
by employing a smaller local measurement SNR $S_M^{(MP)}(V_k){<}S_M^{*}(V_k)$, incurring smaller sensing cost,
 so that the overall sensing-transmission cost is the same for both schemes.
Finally, note that, for both schemes, the local measurement SNR is approximately constant for all values of the quality state $V_k$, thus suggesting
that adaptation of the activation probability is more critical than adaptation of the local measurement SNR. 
A practical implication is that
a lower optimization complexity can be achieved by 
adapting only the former, while
using a constant value for the latter.

\begin{figure}[t]
\centering
\includegraphics[width = .85\linewidth,trim = 10mm 4mm 10mm 9mm,clip=true]{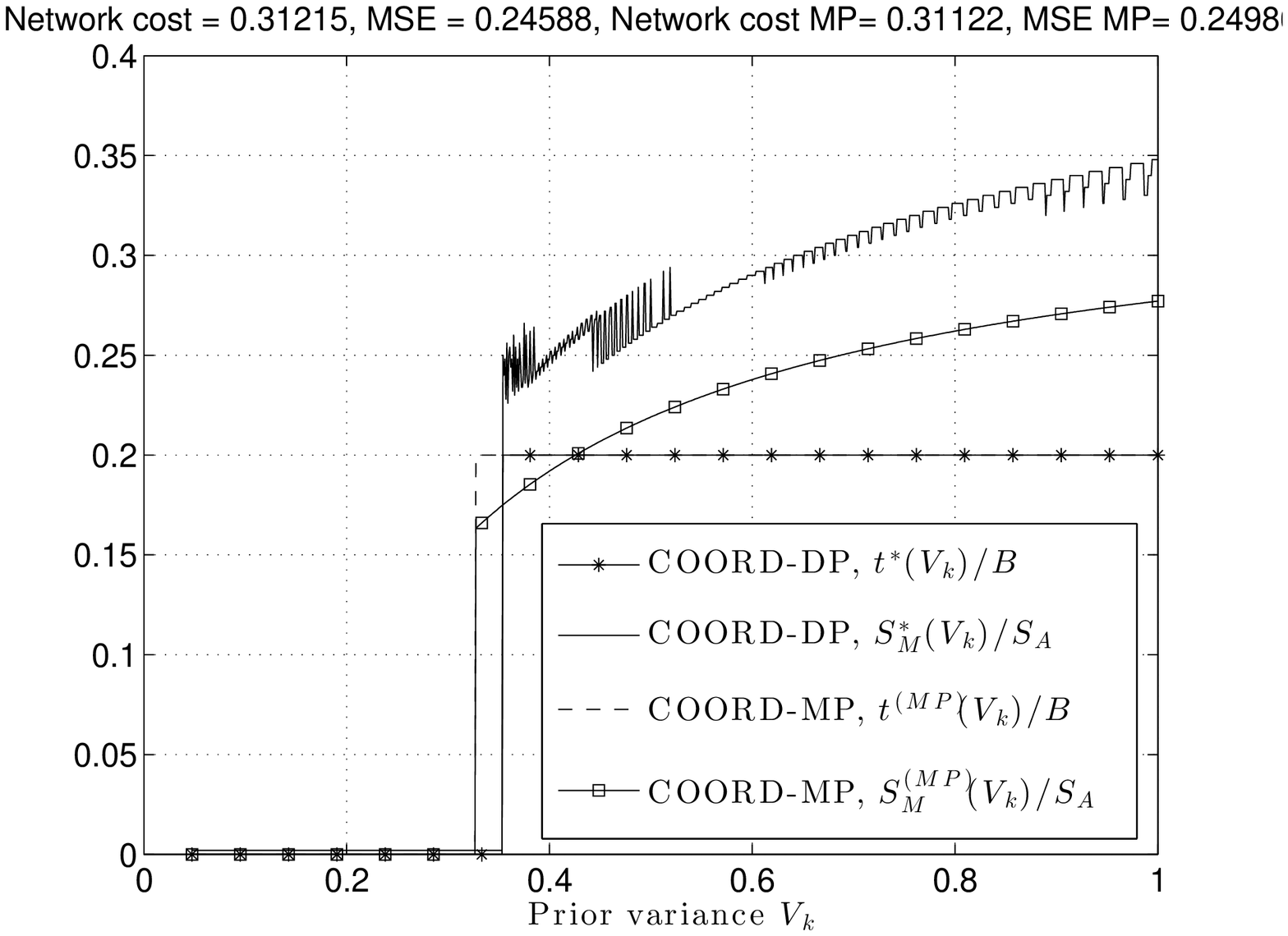}
\vspace{-3mm}
\caption{Structure of COORD-DP and  COORD-MP  as a function of the prior variance $V_k$.
The corresponding simulated network cost is $\simeq 0.312$ and the MSE is $\simeq 0.25$ for both schemes.}
\vspace{-5mm}
\label{figstructCOORD}
\end{figure}

Finally, in Fig. \ref{figstructCOORD}, we plot COORD-DP and COORD-MP as a function of the quality state $V_k$.
Similar to the decentralized scheme, as proved in Theorem~\ref{lemMPclosedform},
  activations are of threshold type, \emph{i.e.}, one SN is activated only if $V_k>0.35$, otherwise all SNs remain idle.
Moreover, as can be observed from the figure and analytically from (\ref{optSM}),
 the local measurement SNR increases with $V_k$, in order to achieve higher estimation accuracy when
the estimation quality at the FC is poor.
\vspace{-3mm}
\section{Conclusions}\label{conclusions}
 \noindent In this paper, we have proposed a cross-layer distributed sensing-estimation framework for WSNs,
 which exploits the quality feedback information from the FC.
 Our cross-layer design approach allows one to model the time-varying capability 
 of the SNs to accurately sense the underlying process,
  the scarce channel access resources shared by the SNs, as well as sensing-transmission costs.
We have proposed a coordinated scheme, where the FC schedules the action of each SN,
 and a more scalable decentralized scheme, where each SN performs a local decision to sense-transmit or remain idle, based on the FC quality feedback and
 the local observation quality.
  In this second part, we have designed low-complexity myopic policies.
For the coordinated scheme, we have shown that the myopic policy can be characterized in closed form. For the decentralized scheme, we have presented an iterative algorithm which converges provably to a local optimum of the myopic cost function.
Numerically, we have shown that the myopic policies achieve near-optimal performance,
 at a fraction of the complexity with respect to the optimal policy derived via dynamic programming,
 and thus are more suitable for implementation in practical WSN deployments.


\appendices
\vspace{-3mm}
\section{}
\label{proofoflemMPclosedform}
\noindent\emph{Proof of Theorem \ref{lemMPclosedform}:}
We first optimize (\ref{MPcost}) with respect to the local measurement SNR $S_M$, for a fixed $t>0$.
Since (\ref{MPcost}) is convex with respect to $S_M$, by computing the derivative with respect to
$S_M$ and setting it to zero, and forcing the solution to be non-negative, since $S_M\geq 0$, we obtain the optimal $S_M^*(t)$
\begin{align}
\label{SMopt}
S_M^*(t)=\left(\frac{1}{\sqrt{\lambda\theta}}-\frac{1}{V_k}\right)^+\frac{S_AV_k}{1+tS_AV_k}.
\end{align}
We now optimize with respect to the number of active SNs $t\in\{0,1,\dots,B\}$.
Note that, if $V_k\leq\sqrt{\lambda\theta}$, then $S_M^*(t)=0,\ \forall t$, hence
 the optimal number of active SNs is $t^{(MP)}(V_k)=0$.
 Otherwise ($V_k>\sqrt{\lambda\theta}$), after plugging $S_M^*(t)$ into the cost function (\ref{MPcost}), we obtain the cost function
\begin{align}
f(t)\triangleq\frac{V_k+2tS_AV_k\sqrt{\lambda\theta}-t\lambda\theta S_A}{1+tS_AV_k}+\lambda t,
\end{align}
hence $t^{(MP)}(V_k)=\arg\min_{t\in\{0,1,\dots,B\}} f(t)$.
In order to solve this problem, we study the function $f(t)$.
We have
\begin{align*}
&g(t)\triangleq (f(t+1)-f(t))(1+(t+1)S_AV_k)(1+tS_AV_k)
\\&=\nonumber
-S_A(\sqrt{\lambda\theta}-V_k)^2
 +\lambda[1+(t+1)S_AV_k](1+tS_AV_k),
\end{align*}
hence $f(t+1)\geq f(t)\Leftrightarrow g(t)\geq 0$.
Note that 
\begin{align*}
g(t+1)-g(t)=2S_AV_k\lambda[1+(t+1)S_AV_k]>0,
\end{align*}
hence $g(t)$ is an increasing function of $t$.
Solving with respect to $V_k$, $g(t)\leq 0$ is equivalent to
\begin{align}
\label{cond1000}
&V_k^2S_A(\lambda(t+1)tS_A-1)
\\&\nonumber
+V_kS_A[2\sqrt{\lambda\theta}+\lambda(2t+1)]
+\lambda (1-S_A\theta)
 \leq 0.
 \end{align}
 Note that (\ref{cond1000}) cannot hold if $[\lambda(t+1)tS_A-1]{\geq}0$, since $V_k{>}\sqrt{\lambda\theta}$ and the left hand expression would be strictly positive. 
 Therefore, $\lambda(t+1)tS_A-1<0$ for (\ref{cond1000}) to hold. Solving with respect to $V_k$, it can be shown that
 (\ref{cond1000}) is equivalent to the union of $V_k\geq v_{\mathrm{th}}(\lambda,t)$ and
  \begin{align}
  \label{sdfsfds}
& V_k\leq 
 \frac{
\sqrt{\lambda\theta}+\frac{\lambda}{2}(2t+1)
 }
 {1-\lambda(t+1)tS_A}
\\& \nonumber
 \!-\!\frac{
 \sqrt{\lambda}
 \sqrt{
 \sqrt{\lambda\theta}(2t\!+\!1)
 \!+\!\lambda \theta(t\!+\!1)tS_A
  \!+\!\frac{\lambda}{4}
 \!+\!\frac{1}{S_A}
 }
 }{1-\lambda(t+1)tS_A}
 \leq\sqrt{\lambda\theta},
 \end{align}
 where the second inequality in (\ref{sdfsfds}) can be proved using the fact that $\lambda<1/[(t+1)tS_A]$ for (\ref{cond1000}) to hold.
 Note that, since $V_k>\sqrt{\lambda\theta}$, the inequality (\ref{sdfsfds}) cannot hold, hence
 \begin{align}
 \label{must}
\!\!\!g(t)\!\leq\!0\!\Leftrightarrow \!
V_k\geq v_{\mathrm{th}}(\lambda,t)\text{ and }\lambda(t+1)tS_A-1<0.
 \end{align}
Let $t^*{=}\max\{t:\lambda(t+1)tS_A-1{<}0\}$, whose solution is given as in the statement of the theorem.
 Clearly, $0{\leq}t^*{<}\infty$.
From (\ref{must}), we then have $g(\tau){>}0,\forall \tau{>}t^*$.
On the other hand, for $\tau\leq t^*$, we have that 
$g(\tau)\leq 0\Leftrightarrow V_k\geq v_{\mathrm{th}}(\lambda,\tau)$.
Note that $v_{\mathrm{th}}(\lambda,\tau){>}v_{\mathrm{th}}(\lambda,\tau-1)$.
It follows that, if $V_k{<}v_{\mathrm{th}}(\lambda,0)$, then 
$V_k{<}v_{\mathrm{th}}(\lambda,\tau),\forall \tau$, and therefore $g(\tau){>}0,\forall \tau$.
In this case, $f(\tau+1){>}f(\tau){>}\dots{>}f(0)$, hence $t^{(MP)}(V_k){=}0$.
On the other hand, if $V_k{\geq}v_{\mathrm{th}}(\lambda,t^*)$, then $V_k{\geq}v_{\mathrm{th}}(\lambda,\tau),\forall\tau{\leq}t^*$,
hence $g(\tau){\leq}0,\forall\tau{\leq}t^*$, $g(\tau){>}0,\forall \tau{>}t^*$. In this case,
$f(t^*{+}1){=}\min_{t}f(t)$, hence $t^{(MP)}(V_k){=}\min\{t^*{+}1,B\}$.
Finally, if
 $v_{\mathrm{th}}(\lambda,t^*){>}V_k{\geq}v_{\mathrm{th}}(\lambda,0)$, 
 letting $\hat t{=}\min\{t{\leq}t^*{:}V_k{<}v_{\mathrm{th}}(\lambda,t)\}$,
we have $V_k{<}v_{\mathrm{th}}(\lambda,\hat t)$, or equivalently $g(\hat t){>}0$,
hence $g(\tau){>}0,\forall \tau{\geq}\hat t$, and
$V_k{\geq}v_{\mathrm{th}}(\lambda,\tau),\forall \tau{<}\hat t$, or equivalently $g(\tau){\leq}0$.
In particular, $g(\hat t{-}1){\leq}0$ and $g(\hat t){>}0$, \emph{i.e.},
$f(\hat t){=}\min_{t\geq 0} f(t)$ and $t^{(MP)}(V_k){=}\min\{\hat t,B\}$.
If $g(\hat t{-}1){=}0$, we have $f(\hat t){=}f(\hat t{-}1)$, hence both $\tau{=}\hat t$ and $\tau{=}\hat t{-}1 $ minimize $f(t)$ and
the choice of $t^{(MP)}(V_k)$ is probabilistic.

To conclude, we show that it suffices to consider $\lambda\leq\lambda_{\mathrm{th}}$.
We show that, if $\lambda>\lambda_{\mathrm{th}}$, then the MP solution is forced to $t^{(MP)}(V_k)=0,\ \forall V_k$, so that all SNs remain idle at all times.
This occurs if $1<v_{\mathrm{th}}(\lambda,0)$, since $V_k\leq 1$, \emph{i.e.},
\begin{align}
\label{aaaaa}
\sqrt{\lambda}\sqrt{\sqrt{\lambda\theta}+\frac{\lambda}{4}+\frac{1}{S_A}}
>
1-\sqrt{\lambda\theta}-\frac{\lambda}{2},
 \end{align}
or equivalently:

\noindent 1) If the right hand expression in (\ref{aaaaa}) is negative, \emph{i.e.},
$\lambda{>}\frac{4}{(\sqrt{\theta+2}+\sqrt{\theta})^2}$;

\noindent 2) If $
\lambda\leq\frac{4}{(\sqrt{\theta+2}+\sqrt{\theta})^2}
$ and, by squaring each side of (\ref{aaaaa}),
\begin{align}
\label{rrr}
\lambda(1-\theta+1/S_A)+2\sqrt{\lambda\theta}-1>0.
\end{align}
We further distinguish the following subcases:

\noindent 2.a) if $\theta=1+1/S_A$, then (\ref{rrr}) is equivalent to $\lambda>\frac{1}{4\theta}$;

\noindent 2.b) if $\theta<1+1/S_A$, then (\ref{rrr}) is equivalent to $\lambda>\lambda_{\mathrm{th}}$;

\noindent 2.c) finally, if $\theta>1+1/S_A$, then (\ref{rrr}) is equivalent to
\begin{align*}
\label{}
\!\!\lambda_{\mathrm{th}}
\!<\!\lambda\!<\!
\frac{1}{(\sqrt{1+1/S_A}-\sqrt{\theta})^2}\!.\!\!
\end{align*}
Note that the upper bound is redundant since, 
using the fact that 
$\theta>1+1/S_A$, we obtain the tighter bound
\begin{align*}
\label{}
&\lambda\leq\frac{4}{(\sqrt{\theta\!+\!2}\!+\!\sqrt{\theta})^2}
\!\!<\!\!
\frac{1}{(\sqrt{1\!+\!1/S_A}\!-\!\sqrt{\theta})^2},
\end{align*}
hence (\ref{rrr}) is equivalent to $\lambda>\lambda_{\mathrm{th}}.$

Combining the cases 1) and 2), (\ref{aaaaa}) holds  if $\lambda{>}\lambda_{\mathrm{th}}$.
Hence, in order to avoid the trivial MP solution $t^{(MP)}(V_k){=}0,\forall V_k$,
 $\lambda$ must satisfy the condition of the theorem.
 
 Finally, the optimal $S_M^{(MP)}(V_k)$ is given by $S_M^{(MP)}(V_k)=S_M^*(t^{(MP)}(V_k))$. The theorem is thus proved.
\hfill\QED

\vspace{-3mm}
\section{}
\label{proofoflemperformance}
\noindent\emph{Proof of Prop. \ref{lemperformance}:}
When $\lambda\to 0$, we have
$v_{\mathrm{th}}(0,t){=}0,\forall t{\geq}-1$,
$t^*{\to}\infty$.
Therefore, since $V_k{>}v_{\mathrm{th}}(0,t),\forall t{\geq}-1$, from Theorem~\ref{lemMPclosedform} we have $t^{(MP)}(V_k){=}B$, hence all channels are occupied.
Moreover, $S_M^{(MP)}(V_k){\to}\infty$, so that the sensing-transmission cost in each slot is $\infty$, and the aggregate SNR collected at the FC in each slot is $\Lambda_k{\to} BS_A,\forall k{\geq}0$. The result follows from \cite[Prop.~7]{MichelusiP1}.
Now, consider the case $\lambda{\to}\lambda_{\mathrm{th}}$. In this case, we have $v_{\mathrm{th}}(\lambda_{\mathrm{th}},0){=}1$,
by definition of $\lambda_{\mathrm{th}}$. Therefore, it follows that $t^{(MP)}(V_k){=}0$, so that
 the sensing-transmission cost in each slot $0$, and the aggregate SNR collected at the FC in each slot is $\Lambda_k{=}0$.
\hfill\QED

\vspace{-3mm}
\section{}
\label{proofofeta}
\begin{propos}[Properties of $\eta_j(\lambda)$ and $\lambda_j^*$]
\label{eta}
$\eta_j(\lambda)$ is a decreasing function of $\lambda{\in}[0,\lambda_{\mathrm{th}}]$ and increasing function of $j{\geq}0$.
Additionally, $\lambda_0^*{=}0$, $\lambda_{j-1}^*{<}\lambda_j^*,\forall j{\geq}1$, and $\lambda_{\infty}^*{\triangleq}\lim_{j\to\infty}\lambda_j^*{=}\lambda_{\mathrm{th}}$.
\end{propos}
\noindent\emph{Proof:}
The first part can be proved by inspection, \emph{i.e.}, by solving  $\frac{\mathrm d \eta_j(\lambda)}{\mathrm d\lambda}{<}0$ and
$\eta_{j+1}(\lambda){-}\eta_j(\lambda){>}0$.
We have $\eta_0(0){=}0$, hence $\lambda_0^*{=}0$, $\lim_{j\to\infty}\eta_j(\lambda){=}1{-}v_{\mathrm{th}}(\lambda,0)$,
and $\lambda_{\infty}^*{=}\lambda_{\mathrm{th}}$.
Finally, $0{=}\eta_{j-1}(\lambda_{j-1}^*){<}\eta_{j}(\lambda_{j-1}^*)$, and thus necessarily $\lambda_{j}^*{>}\lambda_{j-1}^*$, since  $\eta_{j}(\lambda)$
is a decreasing function of $\lambda$.~\hfill\QED

\vspace{-3mm}
\section{}
\label{proofoflemSAinf}
\noindent\emph{Proof of Theorem~\ref{lemSAinf}:}
We prove the theorem only for the case $\lambda{<}\lambda_{\mathrm{th}}$ and $V_0{=}1$. A similar proof holds for the case
$\lambda{=}\lambda_{\mathrm{th}}$ or $V_0{<}1$, the only difference being in the initial transient behavior (which does not affect the
average long-term performance). In the proof, we define $f_i\triangleq 1-\alpha^i(1-\sqrt{\lambda\theta})$, for $i\geq 0$.

Let $\lambda{\in}(\lambda_{J-1}^*,\lambda_J^*]$, for some $J\geq 1$ (for any $\lambda{<}\lambda_{\mathrm{th}}$, such $J$ exists and is unique).
Since $\lambda{<}\lambda_{\mathrm{th}}$, we have $v_{\mathrm{th}}(\lambda,0){<}1{=}V_0$, hence,
from Corollary \ref{corolMPclosedform}, $t^{(MP)}(V_0){=}1$, $\Lambda_0{=}\frac{1}{\sqrt{\lambda\theta}}{-}1$.
Then we have $\hat V_0{=}\sqrt{\lambda\theta}$,  $V_1{=}f_1$,
with cost $c_{\mathrm{TX}}+\phi(1/\sqrt{\lambda\theta}-1)$.

In the following stages $k\geq 1$, let $V_k=f_i$ for some $i>0$. This is true for $k=1$, since $V_1=f_1$.
Then, from Corollary~\ref{corolMPclosedform}:

\noindent 1) if $f_i{<}v_{\mathrm{th}}(\lambda,0)$, then $t^{(MP)}(V_k){=}0$, $\Lambda_k{=}0$, $\hat V_k{=}f_i$,
$V_{k+1}{=}1{-}\alpha(1{-}V_k){=}f_{i+1}$, with cost $0$;

\noindent 2) if $f_i{=}v_{\mathrm{th}}(\lambda,0)$, then, with probability $(1{-}p_0)$, $t^{(MP)}(V_k){=}0$, $\Lambda_k{=}0$, $\hat V_k{=}f_i$,
$V_{k+1}{=}1{-}\alpha(1{-}\hat V_k){=}f_{i+1}$, with cost $0$;
otherwise, with probability $p_0$,
$t^{(MP)}(V_k){=}1$, $\Lambda_k{=}\frac{1}{\sqrt{\lambda\theta}}{-}\frac{1}{f_i}$, $\hat V_k{=}\sqrt{\lambda\theta}$,
$V_{k+1}{=}1{-}\alpha(1{-}\hat V_k){=}f_{1}$, with cost $c_{\mathrm{TX}}{+}\phi(1/\sqrt{\lambda\theta}-f_i^{-1})$;
  
  \noindent 3) if $f_i{>}v_{\mathrm{th}}(\lambda,0)$, then
$t^{(MP)}(V_k){=}1$, $\Lambda_k{=}\frac{1}{\sqrt{\lambda\theta}}{-}\frac{1}{f_i}$, $\hat V_k{=}\sqrt{\lambda\theta}$,
$V_{k+1}{=}f_{1}$, with cost $c_{\mathrm{TX}}{+}\phi(1{/}\sqrt{\lambda\theta}{-}f_i^{-1})$.

Since $\{f_i,i>0\}$ is a non-decreasing sequence,
and using the definition of $\lambda_j^*$ as the unique solution of $\eta_j(\lambda_j^*)=0$ (see (\ref{etak})),
 we have that $f_{i}<v_{\mathrm{th}}(\lambda,0)\Leftrightarrow i<J$,
 and $f_{i}=v_{\mathrm{th}}(\lambda,0)\Leftrightarrow \lambda=\lambda_{J}^*$ and  $i=J$.
It follows that, if $V_k=f_i$ for some $i< J$, then $V_{k+j}=f_{i+j},\forall j\leq  J-i$.
If $V_k=f_{J}$, then, with probability $p_0$ (where $p_0=1$ if $\lambda\in(\lambda_{J-1}^*,\lambda_{J}^*)$), $V_{k+1}=f_1$; otherwise, $V_{k+1}=f_{J+1}$.
Finally, if $V_k=f_{J}+1$, then $V_{k+1}=f_1$.
The prior variance process $\{V_k,k>0\}$ thus follows a time-homogeneous, finite-state Markov chain,
taking value from the set $\{f_1,f_2,\dots,f_{J+1}\}$. 
 Let $\pi_i$ be the long-term time-average probability that $V_k=f_i$, defined as 
\begin{align}
\label{}
\pi_i=\lim_{T\to\infty}\frac{1}{T+1}\sum_{k=0}^T\chi(V_k=f_i).
\end{align}
By solving the steady state equations, it is given by
\begin{align}
\label{}
\pi_i=\begin{cases}
\frac{1}{J+1-p_0}      & i=1,2,\dots,J, \\
\frac{1-p_0}{J+1-p_0}          & i=J+1, \\
0          & i>J+1.
\end{cases}
\end{align}
By averaging with respect the steady-state distribution $\pi_i$,
the average long-term sensing-transmission cost incurred by each SN under the MP is thus given by
\begin{align}
\label{}
&\bar C_{MP}=
\frac{1}{N_S}\pi_{J}p_0\left[c_{\mathrm{TX}}+\phi\left(\frac{1}{\sqrt{\lambda\theta}}-\frac{1}{f_{J}}\right)\right]
\nonumber\\&
+\frac{1}{N_S}\pi_{J+1}\left[c_{\mathrm{TX}}+\phi\left(\frac{1}{\sqrt{\lambda\theta}}-\frac{1}{f_{J+1}}\right)\right],
\end{align}
since transmissions occur only if $V_k=f_{J}$ (with probability $p_0$) or 
$V_k=f_{J+1}$ (with probability $1$), yielding (\ref{Cinf}).
Similarly, the average long-term MSE is given by
\begin{align*}
\label{}
&\bar M_{MP}\!\!=\!\!\!\sum_{i=1}^{J-1}\!\!\pi_if_i\!+\!\pi_{ K_\lambda}(p_0\sqrt{\lambda\theta}\!+\!(1\!-\!p_0)f_{J})
\!+\!\pi_{J+1}\sqrt{\lambda\theta},
\end{align*}
since no transmissions occur in states $f_i,\ i=1,2,\dots,J-1$, hence $\hat V_k=V_k=f_i$,
yielding  (\ref{Rinf}).
\hfill\QED

\vspace{-3mm}
\section{}
\label{proofoflemconvex}
\noindent\emph{Proof of Prop. \ref{lemconvex}:}
Using the fact that $p_0{=}1$ for $\lambda{\in}(\lambda_{j-1}^*,\lambda_j^*)$, we
obtain that
the average long-term expressions (\ref{Rinf}) and (\ref{Cinf}) are continuous functions of 
$\lambda\in(\lambda_{j-1}^*,\lambda_j^*)$.
Similarly, (\ref{Rinf}) and (\ref{Cinf}) are continuous functions of $p_0{\in}[0,1]$, for $\lambda{=}\lambda_j^*,\forall j$.
Continuity at the boundaries holds by inspection of (\ref{Rinf}), (\ref{Cinf}).

Now, we prove that $\bar M_{MP}^{\lambda,1}$ and $\bar C_{MP}^{\lambda,1}$
are, respectively, increasing and decreasing functions of $\lambda\in(\lambda_{j-1}^*,\lambda_j^*),\forall j$,
and that $\bar M_{MP}^{\lambda_j^*,p_0}$ and $\bar C_{MP}^{\lambda_j^*,p_0}$
are, respectively, decreasing and increasing functions of $p_0\in[0,1],\ \forall j$.
The property (\ref{decreasing}) then follows from this and the continuity.
From (\ref{Rinf}) and (\ref{Cinf}), for $j\geq 1$ and $\lambda\in(\lambda_{j-1}^*,\lambda_j^*)$ we have
\begin{align}
\label{derivRinf}
&\frac{\mathrm d\bar M_{MP}^{\lambda,1}}{\mathrm d\lambda}
=
\frac{1-\alpha^{j}}{1-\alpha}\frac{\sqrt{\theta}}{2j\sqrt{\lambda}}>0,
\\&
\label{derivCinf}
\frac{\mathrm d\bar C_{MP}^{\lambda,1}}{\mathrm d\lambda}
\!=\!
\frac{-\phi(1\!-\!\alpha^{j})[1\!-\!\alpha^{j}(1\!-\!\sqrt{\lambda\theta})^2]}{2k\lambda\sqrt{\lambda\theta}[1-\alpha^{j}(1-\sqrt{\lambda\theta})]^2}
<0,
\end{align}
where we have used the fact that $\lambda\leq\lambda_{\mathrm{th}}$, hence $\sqrt{\lambda\theta}\leq 1$.
Similarly, for $j\geq 0$, $\lambda=\lambda_j^*$ and $p_0\in[0,1]$, we have
\begin{align}
\label{}
\label{derivRinfp}
&\frac{\mathrm d\bar M_{MP}^{\lambda,p_0}}{\mathrm dp_0}
=
\frac{1-\sqrt{\lambda_j^*\theta}}{(j+1-p_0)^2}\left(j\alpha^{j}-\frac{1-\alpha^{j}}{1-\alpha}\right),
\end{align}
hence $\frac{\mathrm d\bar M_{MP}^{\lambda,p_0}}{\mathrm dp_0}{<}0{\Leftrightarrow}F_j{\triangleq}j\alpha^{j}-\frac{1-\alpha^{j}}{1-\alpha}{<}0$.
This is verified, since
$F_{j+1}{-}F_j{=}{-}(j{+}1)\alpha^{j}(1{-}\alpha){<}0$, so that $F_j{<}F_0{=}0,\forall j{>}0$.
Similarly, 
\begin{align}
\label{derivCinfp}
&\!\!\!\frac{\mathrm d\bar C_{MP}^{\lambda,p_0}}{\mathrm dp_0}
=
 \frac{\phi}{(j+1-p_0)^2}\left(\frac{1}{\theta}+\frac{1}{\sqrt{\lambda_j^*\theta}}\right)
 \\&
\!\!\!-\!\frac{\phi}{(j+1-p_0)^2}\left[\!\!\frac{j+1}{1\!-\!\alpha^{j}(1\!-\!\sqrt{\lambda_j^*\theta})}-\frac{j}{1\!-\!\alpha^{j+1}(1\!-\!\sqrt{\lambda_j^*\theta})}\!\!\right]\!.\!\!
\nonumber
\end{align}
By solving $\eta_j(\lambda_j^*)=0$ by definition of $\lambda_j^*$, 
with respect to $\theta$ as a function of $\lambda_j^*\theta$, and using (\ref{etak}) and (\ref{vth}),
we obtain
\begin{align}
\label{th}
\theta=\frac{\lambda_j^*\theta[1-\alpha^{j}(1-\sqrt{\lambda_j^*\theta})]}{(1-\alpha^{j})^2(1-\sqrt{\lambda_j^*\theta})^2}.
\end{align}
 Replacing (\ref{th}) in (\ref{derivCinfp}), and letting $x{=}\sqrt{\lambda_j^*\theta}{\in}[0,1]$,
 we obtain
\begin{align}
\label{}
&\!\!\!\frac{\mathrm d\bar C_{MP}^{\lambda,p_0}}{\mathrm dp_0}
\propto
1\!-\!\alpha^j
\!-\!\frac{j\alpha^{j}x^2(1-\alpha)}{[1-\alpha^{j}(1-x)][1-\alpha^{j+1}(1-x)]
}
\triangleq G(x),
\nonumber
\end{align}
We have
\begin{align}
\label{}
&\frac{\mathrm d G(x)}{\mathrm d x} 
=
-\frac{j\alpha^{j}x(1-\alpha)}{[1-\alpha^{j}(1-x)]^2[1-\alpha^{j+1}(1-x)]^2}
\\&\nonumber
\times
\left[
x(2-\alpha^{j+1}-\alpha^{j})
+2(1-x)(1-\alpha^{j})(1-\alpha^{j+1})
\right]
\leq 0.
\end{align}
It follows that $G(x)\geq G(1)=1-\alpha^j-j\alpha^{j}(1-\alpha)\geq 0$, hence
 $\frac{\mathrm d\bar C_{MP}^{\lambda,p_0}}{\mathrm d p_0}>~0$, thus proving
 (\ref{decreasing}).
\hfill\QED
\vspace{-3mm}
\section{}
\label{proofoflemMPdistr}
\noindent\emph{Proof of Theorem \ref{lemMPdistr}:}
Let, for $V_k\in(0,1]$,
\begin{align}
\label{}
&S_{M}^{(MP)}(\zeta;V_k)=\arg\min_{S_M\geq 0}f(\zeta,S_M,V_k),\ \zeta>0,
\\
&\zeta^{(MP)}(S_M;V_k)=\arg\min_{\zeta\geq 0}f(\zeta,S_M,V_k),\ S_M\geq 0.
\end{align}
\subsection{Optimal $\zeta^{(MP)}(S_M;V_k)$ given $S_M\geq 0$}
\noindent It can be shown that
\begin{align}
\label{fdsfds}
&\frac{\mathrm df(\zeta,S_M,V_k)}{\mathrm d\zeta}=
\lambda B (1+\theta S_M)
\\&\!+\!\nonumber
e^{-\zeta}(1\!-\!\zeta)
\mathbb E\left[\left.\frac{R_k-\rho(\zeta)B}{\rho(\zeta)(1\!-\!\rho(\zeta))}\hat \nu\left(V_{k},\frac{R_kS_AS_{M}}{S_A+S_{M}}\right)
\right|\rho(\zeta)\right],
\end{align}
where we have used the fact that 
\begin{align}
\frac{\mathrm d\mathcal B_{B}\left(r;\rho\right)}{\mathrm d\rho}=
\mathcal B_{B}\left(r;\rho\right)
\frac{r-\rho B}{\rho(1-\rho)}.
\end{align}
The argument within the expectation in (\ref{fdsfds}) is concave in $R_k$. If $\zeta\geq 1$, using Jensen's inequality \cite{Boyd},
we thus obtain
\begin{align}
&\frac{\mathrm df(\zeta,S_M,V_k)}{\mathrm d\zeta}\geq 
\lambda B (1+\theta S_M)>0,
\end{align}
where we have used the fact that $\mathbb E\left[\left.R_k\right|\rho(\zeta)\right]{=}\rho(\zeta)B$.
It follows that $f(\zeta,S_M,V_k)$ increases for $\zeta{\geq}1$, hence $\zeta^{(MP)}(S_M;V_k){\in}[0,1)$ and we optimize over $\zeta{\in}[0,1) $ hereafter.
By multiplying each side of (\ref{fdsfds}) by $\frac{e^\zeta}{1-\zeta}$, we obtain that
$\frac{\mathrm df(\zeta,S_M,V_k)}{\mathrm d\zeta}{>}0$ is equivalent to $g(S_M,\zeta,V_k){>}0$.
We have the following property of $g(S_M,\zeta,V_k)$.
\begin{propos}
\label{incrG}
$g(S_M,\zeta,V_k)$ is an increasing function of $\zeta$.
\end{propos}
\noindent\emph{Proof:}
See Appendix \ref{AppincrG}.
\hfill\QED

\noindent Using Prop. \ref{incrG} and the fact that 
 $\lim_{\zeta\to 1}g(S_M,\zeta,V_k)=\infty$, we obtain the following cases, depending on the sign of
\begin{align*}
\label{}
g(S_M,0,V_k)\!=\!B\!\left[\!\hat \nu\left(\!V_{k},\frac{S_AS_M}{S_A\!+\!S_M\!}\right)\!-\!V_k\right]
\!+\!\lambda B (1\!+\!\theta S_M)\!:
\end{align*}
  if $g(S_M,0,V_k)\geq 0$, 
  then $\frac{\mathrm df(\zeta,S_M,V_k)}{\mathrm d\zeta}>0,\forall\zeta\in (0,1)$ and $\zeta^{(MP)}(S_M;V_k)=0$;
  otherwise,
 $\zeta^{(MP)}(S_M;V_k)$ is the unique $\zeta\in (0,1)$ such that $g(S_M,\zeta,V_k)=0$.
%
\subsection{Optimal $S_{M}^{(MP)}(\zeta;V_k)$ given $\zeta\in(0,1)$}
\noindent Let $\zeta\in(0,1)$. It can be shown that
\begin{align}
\label{}\nn
&\frac{\mathrm df(\zeta,S_M,V_k)}{\mathrm d S_M}=h(S_M,\zeta,V_k),\ \text{hence}
\\
\label{fderiv}
&\frac{\mathrm d^2f(\zeta,S_M,V_k)}{\mathrm d S_M^2}=
\frac{\mathrm dh(S_M,\zeta,V_k)}{\mathrm d S_M}
\\&\nonumber
\!=\!
\mathbb E\left[\left.\!
\hat \nu\!\left(V_{k},\frac{R_kS_AS_{M}}{S_A+S_{M}}\right)^{\!\!3}\!
\frac{2R_kS_A^2(1+V_kR_kS_A)}{V_k(S_A+S_M)^3}
\!\right|\rho(\zeta)\!
\right]
>0,
\end{align}
hence $f(\zeta,S_M,V_k)$ is convex in $S_M$, for a fixed $\zeta>0$, $V_k\in(0,1]$.
We have
$\underset{S_M\to\infty}{\lim} h(S_M,\zeta,V_k)=\lambda\theta \zeta B>0$ and
\begin{align}
\label{}
h(0,\zeta,V_k)=
-\zeta e^{-\zeta}BV_k^2+\lambda\theta \zeta B.
\end{align}
Then, if $h(0,\zeta,V_k){\geq}0$, \emph{i.e.},
$\zeta{\geq}\zeta_{\mathrm{th}}^{\max}(V_k)$, we have $h(S_M,\zeta,V_k){\geq}0,\forall S_M{\geq}0$, hence 
$S_{M}^{(MP)}(\zeta;V_k){=}0$.
Otherwise ($\zeta{<}\zeta_{\mathrm{th}}^{\max}(V_k)$),
$S_{M}^{(MP)}(\zeta;V_k)$ is the unique $S_M{>}0$ such that $h(S_M,\zeta,V_k){=}0$.
By evaluating $h(S_M,\zeta,V_k)$ in $S_M{=}S_A\left(\frac{V_k}{\sqrt{\lambda\theta}}{-}1\right)$, it can be shown that
\begin{align}
\label{}
h\left(S_A\left(V_k/\sqrt{\lambda\theta}-1\right),\zeta,V_k\right)>0.
\end{align}
Therefore, necessarily $S_{M}^{(MP)}(\zeta;V_k)\in (0,S_A\frac{V_k-\sqrt{\lambda\theta}}{\sqrt{\lambda\theta}})$.

We now prove that the MP is $\zeta^{(MP)}(V_k){=}0{\Leftrightarrow}V_k{\leq}v_{\mathrm{th}}(\lambda,0)$.
In fact, if there exists some $\tilde S_M\geq 0$ such that $g(\tilde S_M,0,V_k){<}0$, for such $\tilde S_M$ we have that 
$\zeta^{(MP)}(\tilde S_M;V_k){>}0$
and, for all $S_M{\geq}0$,
$V_k{=}f(0,S_M,V_k){>}f(\zeta^{(MP)}(\tilde S_M;V_k),\tilde S_M,V_k)$, 
hence the MP satisfies $\zeta^{(MP)}{>}0$ (in fact, $\zeta{=}0$ has sub-optimal cost $f(0,S_M,V_k){=}V_k$).

On the other hand, if $g(S_M,0,V_k){\geq}0,\forall S_M{\geq}0$, it follows that
$\zeta^{(MP)}(S_M;V_k){=}0,\forall S_M{\geq}0$, hence the MP satisfies  $\zeta^{(MP)}(V_k){=}0$.
We conclude that $\zeta^{(MP)}(V_k){=}0{\Leftrightarrow}\min_{S_M\geq 0}g(S_M,0,V_k){\geq}0$.
We thus minimize $g(S_M,0,V_k)$ with respect to $S_M$. It can be shown that $g(S_M,0,V_k)$ is a convex function of $S_M{\geq}0$.
By setting the derivative with respect to $S_M$ to zero and forcing the solution to be non-negative (since $S_M{\geq}0$), we obtain
\begin{align}
S_M^*=\left(\frac{1}{\sqrt{\lambda\theta}}-\frac{1}{V_k}\right)^+\frac{S_AV_k}{1+S_AV_k}.
\end{align}
By evaluating the function $g(S_M^*,0,V_k)$ when $V_k\leq \sqrt{\lambda\theta}$, hence $S_M^*=0$,
we obtain
$g(S_M^*,0,V_k)=\lambda B\geq 0$, hence $\zeta^{(MP)}=0$ if $V_k{\leq}\sqrt{\lambda\theta}$.
We now consider the case $V_k{>}\sqrt{\lambda\theta}$. After rearranging the terms, we obtain
\begin{align*}
\label{}
g(S_M^*,0,V_k)=&B\lambda
-B\frac{S_A}{1+S_AV_k}(V_k-\sqrt{\lambda\theta})^2.
\end{align*}
Solving $g(S_M^*,0,V_k)\geq 0$ with respect to $V_k$, it can be shown that this is equivalent to $V_k\leq v_{\mathrm{th}}(\lambda,0)$,
and therefore $\zeta^{(MP)}(V_k)=0\Leftrightarrow V_k\leq v_{\mathrm{th}}(\lambda,0)$.

Finally, we  show that the MP lies within (\ref{boundz}) and (\ref{boundSM}), when $V_k{>}v_{\mathrm{th}}(\lambda,0)$.
By contradiction, if $\zeta^{(MP)}(V_k){\geq}\zeta_{\mathrm{th}}^{\max}(V_k)$,
then $S_{M}^{(MP)}(V_k){=}0$, hence
$\zeta^{(MP)}(V_k){=}\arg\min f(\zeta,0,V_k){=}0$, yielding a contradiction.
Hence, necessarily,  $\zeta^{(MP)}(V_k){<}\zeta_{\mathrm{th}}^{\max}(V_k)$.
On the other hand, if $g(S_{M}^{(MP)}(V_k),0,V_k){\geq}0$, then $\zeta^{(MP)}(V_k){=}0$, yielding a contradiction.
Therefore, we must have $g(S_{M}^{(MP)}(V_k),0,V_k){<}0$. By solving it with respect to 
$S_{M}^{(MP)}(V_k)$, we obtain (\ref{boundSM}).
Using the fact that $V_k{>}v_{\mathrm{th}}(\lambda,0)$, it can be shown  that 
$S_{M,\mathrm{th}}^{\min}{>}0$. Moreover, in general, $S_{M,\mathrm{th}}^{\max}{>}S_{M,\mathrm{th}}^{\min}$,
so that
the upper/lower bounds are not tight.
\hfill\QED
\vspace{-3mm}
\section{}
\label{AppincrG}
\noindent\emph{Proof of Prop. \ref{incrG}:}
We have
 \begin{align*}
\label{}
&\!\frac{\mathrm dg(S_M,\zeta,V_k)}{\mathrm d\zeta}\!=\!
\frac{Be^{-\zeta}(1-\zeta)}{\rho(\zeta)^2(1-\rho(\zeta))^2}
\mathbb E\left[
\hat \nu\left(V_{k},\frac{R_kS_AS_M}{S_A+S_M}\right)\right.
\\&\left.\left.\!\times\!
\vphantom{\hat \nu\left(V_{k},\frac{R_kS_AS_M}{S_A+S_M}\right)}
\left[
(R_k\!-\!\rho(\zeta)B)^2
\!-\!R_k(1\!-\!\rho(\zeta))
\!+\!(R_k\!-\!\rho(\zeta)B)\rho(\zeta)
\right]\right|\rho(\zeta)\right]
\nonumber\\&
+e^{\zeta}\frac{2-\zeta}{(1-\zeta)^2}
\lambda B^2 (1+\theta S_M)\\
&>
B e^{-\zeta}(1\!-\!\zeta)(1\!-\!\rho(\zeta))^{B-2}
s_B\left(\frac{\rho(\zeta)}{1\!-\!\rho(\zeta)},\frac{S_AS_M}{S_A\!+\!S_M}\right),
\end{align*}
where the inequality is obtained by minimizing with respect to $\lambda$, yielding $\lambda=0$, and we have defined, for
$x\in\left[0,\frac{1}{e-1}\right]$ and $S_T\geq 0$,
\begin{align}
\label{sb}
&s_B(x,S_T)=\frac{1}{x^2 }\sum_{r=0}^B
\left(
\begin{array}{c}
B\\r
\end{array}
\right)x^r
\hat \nu\left(V_{k},rS_T\right)
\\&\times
\left[
\left(r(1+x)-xB\right)^2
-r(1+x)
+\left(r(1+x)-xB\right)x
\right].\nonumber
\end{align}
By rearranging the terms, we obtain, for $B>1$.
\begin{align}
\label{}
&s_B(x,\!S_T)
\!=\!\nonumber
B(B\!-\!1)(1\!+\!x)^2
\!\sum_{r=0}^{B\!-\!2}\!\!
\left(
\begin{array}{c}
\!\!\!B\!-\!2\!\!\!\\r
\end{array}
\right)x^{r}
\hat \nu\left(V_{k},(r+2)S_T\right)\\&
-2B(B\!-\!1)(1+x)\!\sum_{r=0}^{B-1}\!\!
\left(
\begin{array}{c}
\!\!\!B-1\!\!\!\\r
\end{array}
\right)x^{r}\nonumber
\hat \nu\left(V_{k},(r+1)S_T\right)\\&
+B(B\!-\!1)\sum_{r=0}^B
\left(
\begin{array}{c}
B\\r
\end{array}
\right)x^r
\hat \nu\left(V_{k},rS_T\right).
\end{align}

We now prove that $s_B(x,S_T)\geq 0$, by induction on $B$.
For $B=1$,  from (\ref{sb}) we obtain $s_1(x,S_T)=0$.
Now, assume that, for some  $B>1$, $s_{B-1}(x,S_T)\geq 0$. We prove that this implies $s_B(x,S_T)\geq 0$.
It can be shown that 
the derivative of $s_B(x,S_T)$ with respect to $x$ is given by
\begin{align}
\label{}
&\!\!\frac{\mathrm ds_B(x,S_T)}{\mathrm dx}\!=\!
\frac{1}{1+V_kS_T}Bs_{B-1}\left(x,\frac{S_T}{1+V_kS_T}\right)\geq 0,
\end{align}
hence $s_B(x,S_T)\geq s_B(0,S_T)$.
The result follows since $s_B(0,S_T)>0$ by inspection.
\hfill\QED
\bibliographystyle{IEEEtranS}
\bibliography{IEEEabrv,References}

\end{document}